\documentclass[twocolumn,superscriptaddress,floatfix,showpacs,aps,longbibliography,10pt]{revtex4-1}
\usepackage[english]{babel}
\usepackage{graphicx}
\usepackage{amsmath}
\usepackage{amssymb}
\usepackage[colorlinks=true, citecolor=blue, urlcolor=blue]{hyperref}
\usepackage{bm}
\usepackage{color}

\DeclareMathOperator{\e}{e}
\DeclareMathOperator{\de}{d\!}
\DeclareMathOperator{\diag}{diag}
\newcommand{\ket}[1]{|#1\rangle}
\newcommand{\bra}[1]{\langle#1|}
\newcommand{\abs}[1]{\lvert#1\rvert} 

\usepackage{color}

\begin{document}

\title{Flux-controlled quantum computation with Majorana fermions}
\author{T. Hyart}
\affiliation{Instituut-Lorentz, Universiteit Leiden, P.O. Box 9506, 2300 RA Leiden, The Netherlands}
\author{B. van Heck}
\affiliation{Instituut-Lorentz, Universiteit Leiden, P.O. Box 9506, 2300 RA Leiden, The Netherlands}
\author{I. C. Fulga}
\affiliation{Instituut-Lorentz, Universiteit Leiden, P.O. Box 9506, 2300 RA Leiden, The Netherlands}
\author{M. Burrello}
\affiliation{Instituut-Lorentz, Universiteit Leiden, P.O. Box 9506, 2300 RA Leiden, The Netherlands}
\author{A. R. Akhmerov}
\affiliation{Department of Physics, Harvard University, Cambridge, Massachusetts 02138 USA}
\author{C. W. J. Beenakker}
\affiliation{Instituut-Lorentz, Universiteit Leiden, P.O. Box 9506, 2300 RA Leiden, The Netherlands}

\date{April 2013}
\begin{abstract}
Majorana fermions hold promise for quantum computation, because their non-Abelian braiding statistics allows for topologically protected operations on quantum information. Topological qubits can be constructed from pairs of well-separated Majoranas in networks of nanowires. The coupling to a superconducting charge qubit in a transmission line resonator (transmon) permits braiding of Majoranas by external variation of magnetic fluxes. We show that readout operations can also be fully flux-controlled, without requiring microscopic control over tunnel couplings. We identify the minimal circuit that can perform the initialization--braiding--measurement steps required to demonstrate non-Abelian statistics. We introduce the Random Access Majorana Memory, a scalable circuit that can perform a joint parity measurement on Majoranas belonging to a selection of topological qubits. Such multi-qubit measurements allow for the efficient creation of highly entangled states and simplify quantum error correction protocols 
by avoiding the need for ancilla qubits.
\end{abstract}
\maketitle

After the first signatures were reported \cite{mourik2012,deng2012,rokhinson2012,das2012} of Majorana bound states in superconducting nanowires \cite{kitaev2001,lutchyn2010,oreg2010}, the quest for non-Abelian braiding statistics \cite{moore1991,read2000, ivanov2001, alicea2011} has intensified. Much interest towards Majorana fermions arises from their technological potential in fault-tolerant quantum computation 
\cite{kitaev2006,das-sarma2005,stern2006,bonderson2006,nayak2008}. Their non-Abelian exchange statistics would allow to perform quantum gates belonging to the Clifford group with extremely good accuracy. Moreover, topological qubits encoded non-locally in well-separated Majorana bound states would be resilient against many sources of decoherence. Even without the applications in quantum information processing, observing a new type of quantum statistics would be a milestone in the history of physics.

The two central issues for the application of Majorana fermions are (i) how to unambiguously demonstrate their non-Abelian exchange statistics and (ii) how to exploit their full potential for quantum information processing. The first issue requires an elementary circuit that can perform three tasks: initialization of a qubit, 
braiding (exchange) of two Majoranas, and finally measurement (readout) of the qubit. 
In view of the second issue, this circuit should be scalable and serve as a first step towards universal fault-tolerant quantum computation.

\begin{figure}[tb]
\centerline{\includegraphics[width=\linewidth]{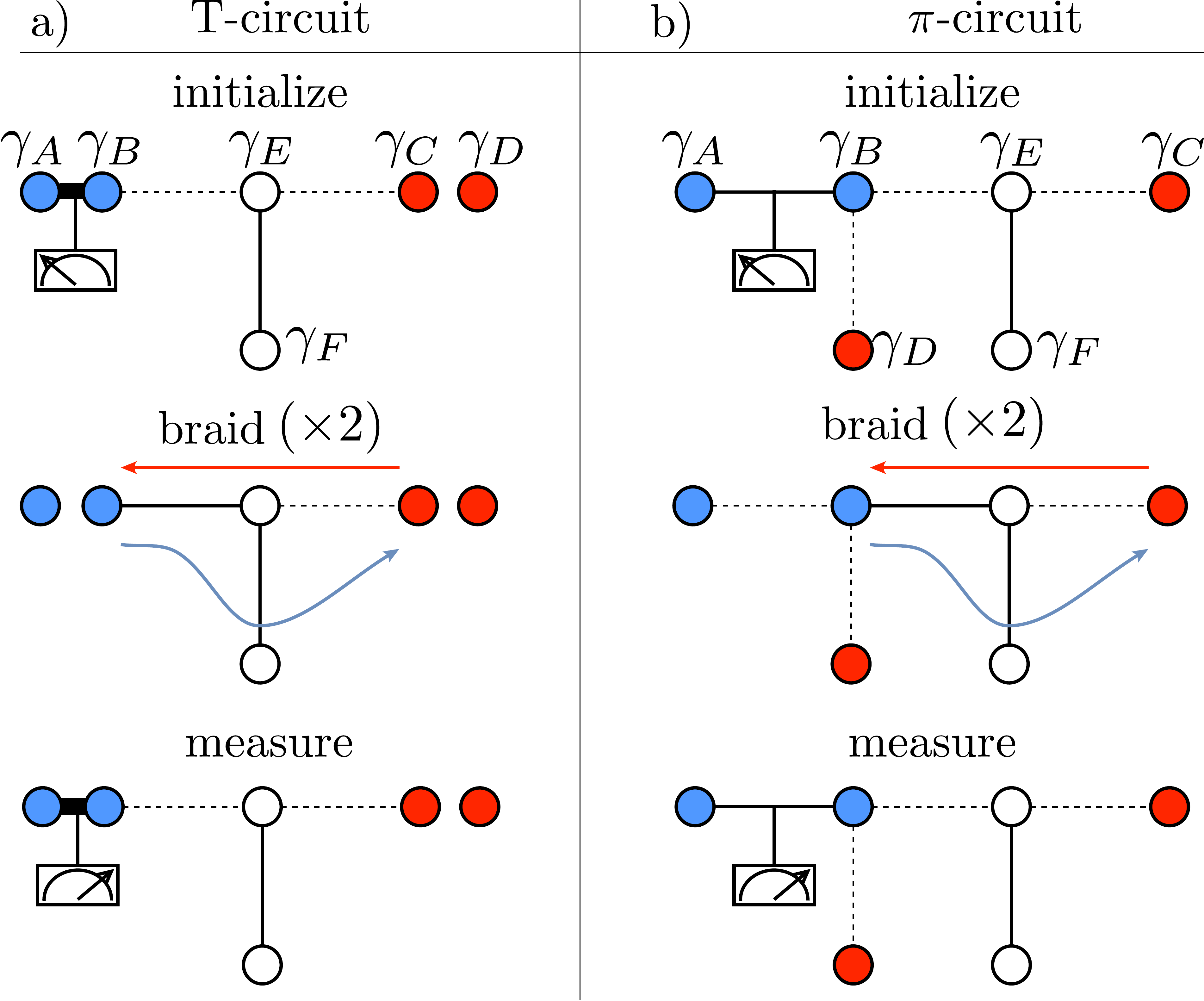}}
\caption{Two circuits that can demonstrate non-Abelian statistics, by the initialization, braiding, and measurement of pairs of Majorana bound states (circles). Braiding is performed twice to flip the fermion parity of $\gamma_A$ and $\gamma_B$ \cite{das-sarma2005}. Majoranas that can be coupled by Coulomb charging energy are connected by a thin line; the line is solid if the Majoranas are strongly coupled, and dashed if they are uncoupled. A thick line indicates tunnel coupling of Majoranas. The T-shaped circuit of Ref.\ \cite{alicea2011} (left column) requires control over tunnel couplings, while the $\pi$-shaped circuit considered here (right column) does not, because both readout \textit{and} braiding involve a Majorana localized at a T-junction.
}
\label{fig:TvsPi}
\end{figure}

Here we present such a circuit, using a superconducting charge qubit in a transmission line resonator (\textit{transmon} \cite{koch2007,schreier2008,dicarlo2009,houck2009}) to initialize, control, and measure the topological qubit. In such a hybrid system, named \textit{top-transmon} \cite{hassler2011}, the long-range Coulomb couplings of Majorana fermions can be used to braid them and to read out their fermion parity \cite{hassler2011, heck2012}.
While there exist several proposals to control or measure Majorana fermions in nanowires \cite{hassler2011, hassler2010, romito2012, heck2012, alicea2011, sau2010, jiang2011, bonderson2011, sau2011, flensberg2011, leijnse2011, halperin2012, bruder}, 
combining braiding \textit{and} measurement without local adjustment of microscopic parameters remains a challenge. We show that full macroscopic control is possible if during the measurement one of the Majorana fermions is localized at a T-junction between three superconducting islands (see Fig.~\ref{fig:TvsPi}).
All three steps of the braiding protocol, initialization--braiding--measurement, can then be performed by adjusting magnetic fluxes through split Josephson junctions. 
Because local control of microscopic parameters is not necessary, our scheme is less sensitive to problems arising from electrostatic disorder and screening of gate voltages by the superconductor.

\begin{figure*}[tb]
\centerline{\includegraphics[width=0.95\textwidth]{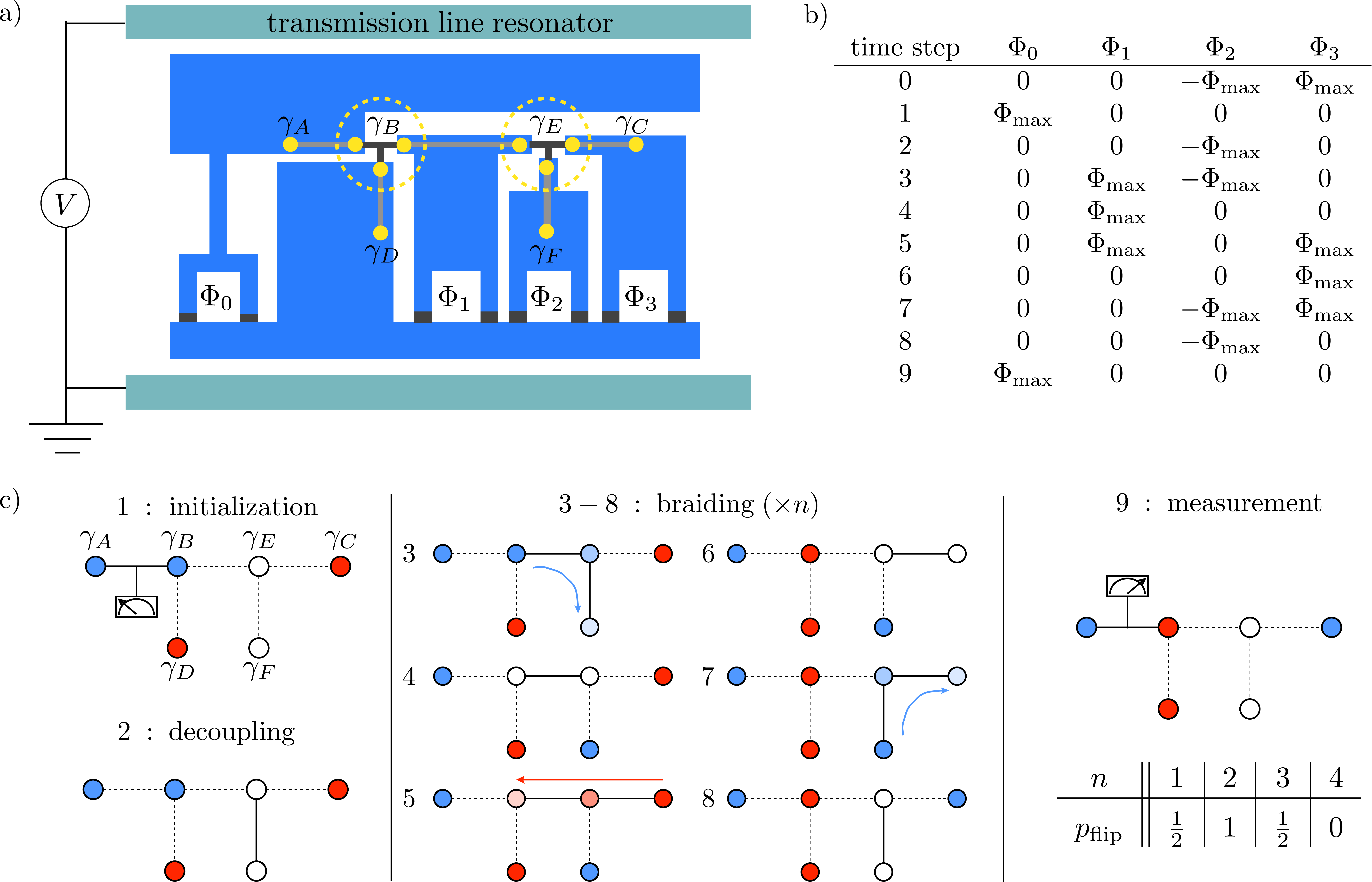}}
\caption{Panel a): Minimal circuit for flux-controlled demonstration of non-Abelian Majorana statistics. Two large superconducting plates form a Cooper pair box in a transmission line resonator, i.e. a transmon qubit. Three smaller superconducting islands are embedded between the two transmon plates. Each superconducting island contains a nanowire supporting two Majorana bound states. At low energies, the three overlapping Majorana bound states at a T-junction form a single zero mode so that effectively the system hosts six Majorana bound states, labeled $\gamma_A$, $\gamma_B$, $\gamma_C$, $\gamma_D$, $\gamma_E$, and $\gamma_F$. The Coulomb couplings between the Majorana fermions can be controlled with magnetic fluxes $\Phi_k$. This hybrid device can measure the result of the braiding operation as a shift in the microwave resonance frequency when the fermion parity $i\gamma_A \gamma_B$ switches between even and odd. Panel b): Sequence of variation of  fluxes during the initialization (steps 0--2), braiding (steps 3--8) and measurement (step 9). Panel c): Illustration of the steps required for initialization, braiding and measurement. Fusion channels of pairs of Majorana fermions colored red, blue and white are chosen to be the basis states in Eq. \eqref{eq:fermionic_basis}.
To unambiguously demonstrate the non-Abelian nature of Majoranas, one needs to collect statistics of measurement outcomes when the adiabatic cycle describing the braiding operation (steps 3--8)  is repeated $n$ times between initialization and measurement. The probabilities of observing changes in the cavity's resonance frequency, $p_{{\rm flip}}$, for different values of $n$ should obey the predictions summarized in the table. The sequence of probabilities shown in the table repeats itself periodically 
for larger values of $n$.} 
\label{fig_protocol}
\end{figure*}

This design principle of flux-controlled braiding and measurements can be scaled up from a minimal braiding experiment setup to a multi-qubit register that supports a universal set of quantum gates and allows measurement of any product of Pauli matrices belonging to a selection of topological qubits. Multi-qubit parity measurements are a powerful resource in quantum information processing, allowing for the efficient creation of long-range entanglement and direct measurement of stabilizer operators (thus removing the overhead of ancilla qubits in quantum error correction schemes).  Because the data stored in the register can be accessed in any random order, it truly represents a Random Access Majorana Memory.

The structure of the paper is as follows. In Sec.~\ref{sec_braiding} we present the circuit that can demonstrate the non-Abelian Majorana statistics. In Sec.~\ref{sec_RAMM} we take a longer-term perspective and describe the Random Access Majorana Memory, whose potential for quantum computation is discussed in Sec.~\ref{sec_multiqubits}. Finally, we conclude in Sec.~\ref{sec_discussion}. For the benefit of the reader, we include more detailed derivations and discussions in the Appendices.

\section{Minimal circuit for the demonstration of non-Abelian statistics}
\label{sec_braiding}

To demonstrate non-Abelian Majorana statistics one needs to read out the parity of two Majoranas, $\gamma_A$ and $\gamma_B$, and braid one of these Majoranas $\gamma_B$ with another one, $\gamma_C$. We seek a transmon circuit that can combine these operations in a fully flux-controlled way, by acting on the Coulomb coupling of the Majoranas. Since $\gamma_B$ must be coupled first to one Majorana (for the braiding) and then to another (for the readout), it must be able to contribute to two \textit{different} charging energies. This is possible if $\gamma_B$ is localized at a T-junction between three superconducting islands. 

We thus arrive at the minimal circuit  shown in Fig.~\ref{fig_protocol}a. It consists of five superconducting islands, each containing a nanowire supporting two Majorana bound states, enclosed in a transmission line resonator. The two bigger superconductors form a transmon qubit and the three smaller islands are embedded between the two transmon plates. The Josephson couplings between the islands can be controlled by magnetic fluxes $\Phi_k$ ($k=0,1,2,3$). The nanowires form a $\pi$-shaped circuit, with two T-junctions where three Majorana bound states belonging to adjacent superconductors are tunnel-coupled. At low energies the three overlapping Majorana bound states at a T-junction form a single zero mode, so that effectively the system hosts six Majorana bound 
states, $\gamma_A,\gamma_B,...,\gamma_F$.

The three relevant energy scales for the device are (i) the charging energy $E_{{\rm C},k}=e^2/2C_k$ determined by the total capacitance $C_k$ of the four upper superconductors in Fig.~\ref{fig_protocol}a, (ii) the Josephson energies $E_{{\rm J},k}(\Phi_{k})=E_{{\rm J},k}(0)\cos(e\Phi_k/\hbar)$, and (iii) the Majorana tunnel couplings $E_{\rm M}$ at both T-junctions.  For strong Josephson coupling, $E_{{\rm J},k} \gg E_{{\rm C},k},E_{\rm M}$, the phases of the order parameter on superconducting islands (measured with respect to the lower superconductor) are pinned to the value $\phi_{k}\equiv 0$. We distinguish two different operating regimes of the device: one for the braiding procedure and one for initialization and readout. 

\textbf{Flux-controlled braiding.} During the braiding procedure we set $\Phi_0=0$ so that the charging energy of the large island can be completely neglected.
The charging energies of the small islands can be considered perturbatively \cite{koch2007}, resulting in long-range Coulomb couplings,
\begin{equation}
U_k = 16\left(\frac{E_{{\rm C},k} E_{{\rm J},k}^3}{2\pi^2}\right)^{\tfrac{1}{4}}\e^{-\sqrt{8E_{{\rm J},k}/E_{{\rm C},k}}} \cos(q_k\pi/e), \label{CoulombCoupApp}
\end{equation}
between the Majorana bound states in the corresponding island  \cite{hassler2011}. The offset charge $q_k$ accounts for the effect of nearby gate electrodes. In order to keep our analytic calculations more transparent, we assume that $U_k\ll E_{\rm M}$. This condition is not required for braiding to stay accurate in view of the topological nature of the latter (see also App.~\ref{app_energies}). In this case,  the low-energy sector of the system is described by the effective Hamiltonian (see Appendix~\ref{app_pi_circ})
\begin{equation}
H_{\rm braiding}=-i\Delta_1\gamma_B\gamma_E-i\Delta_2 \gamma_E\gamma_F -i\Delta_3\gamma_E\gamma_C,
\end{equation}
\begin{subequations}
\begin{align}
\Delta_1 ={}& \frac{U_1}{\sqrt{1+2 \cos^2(e\Phi_1/2\hbar)}}  \nonumber \\
  & \times \frac{\cos\,\alpha_{23}}{\sqrt{\cos^2\alpha_{12}+\cos^2\alpha_{23}+\cos^2\alpha_{31}}},\\
\Delta_2 ={}& U_2\frac{\cos\,\alpha_{31}}{\sqrt{\cos^2\alpha_{12}+\cos^2\alpha_{23}+\cos^2\alpha_{31}}},\\
\Delta_3 ={}& U_3\frac{\cos\,\alpha_{12}}{\sqrt{\cos^2\alpha_{12}+\cos^2\alpha_{23}+\cos^2\alpha_{31}}},
\end{align}
\end{subequations} 
where $\alpha_{12}=(e/2\hbar)(\Phi_1+\Phi_2)$, $\alpha_{23}=(e/2\hbar)(\Phi_2+\Phi_3)$, and $\alpha_{31}=-\alpha_{12}-\alpha_{23}$ are gauge-invariant phase differences between the smaller islands.
The three couplings $\Delta_i$ are all tunable with exponential sensitivity via the fluxes $\Phi_{i}$, increasing from $\Delta_{\rm min}$ (the \textit{off} state) to $\Delta_{\rm max}$ (the \textit{on} state) when $|\Phi_{i}|$ increases from $0$ to $\Phi_{\rm max}<h/4e$. On the other hand, the tunnel couplings at the T-junction vary slowly with the fluxes, so  the three overlapping Majoranas remain strongly coupled throughout the  operation. 

Out of the six Majorana operators, we define three fermionic creation operators:
\begin{subequations}\label{eq:fermionic_basis}
\begin{align}
c_1^\dag&=\tfrac{1}{2}(\gamma_A+i\gamma_B)\\
c_2^\dag&=\tfrac{1}{2}(\gamma_C+i\gamma_D)\\
c_3^\dag&=\tfrac{1}{2}(\gamma_E+i\gamma_F).
\end{align}
\end{subequations}
We will braid the Majoranas $\gamma_B$ and $\gamma_C$ by using $\gamma_E$ and $\gamma_F$ as ancillas, as  specified in Fig.~\ref{fig_protocol}. At the beginning and at the end, the Majoranas $\gamma_E$ and $\gamma_F$ are strongly coupled ($|\Phi_2|=\Phi_{\rm max}$). If all other couplings are \textit{off} we are left with two degenerate states that define a topological qubit. In the odd-parity sector they are ${1\choose 0} =|10\rangle|0\rangle$ and ${0\choose 1}=|01\rangle|0\rangle$. During the  exchange of Majoranas $\gamma_B$ and $\gamma_C$ the fluxes $\Phi_1$, $\Phi_2$, $\Phi_3$ are varied between $0$ and $\pm\Phi_{\rm max}$ according to the table shown in Fig.~\ref{fig_protocol}b. 
Computing the non-Abelian Berry phase for this adiabatic cycle as in Ref.\ \cite{heck2012} shows that braiding has the effect of multiplying the topological qubit state with the matrix
\begin{equation}
{\cal U}=\frac{1}{\sqrt{2}}\begin{pmatrix}
1&-i\\
-i&1
\end{pmatrix},\label{Udef}
\end{equation}
up to corrections of order $\Delta_{\rm min}/\Delta_{\rm max}$, with $\Delta_{\rm min}/\Delta_{\rm max}\ll 1$ because of the exponential sensitivity of these quantities on magnetic fluxes. Repeating the cycle $n$ times corresponds to applying the gate ${\cal U}^n$.

\textbf{Initialization and readout.} The ancillas need to be initialized in the state $\ket{0}$. This can be achieved by turning the couplings $\Delta_2$ and $\Delta_3$ on and allowing the system to relax to the ground state by adiabatically switching off $\Delta_3$ before $\Delta_2$ [step 0 in Fig.~\ref{fig_protocol} (b)]. In addition to the initialization of the ancillas, the braiding needs to be preceded and followed by a readout of the topological qubit. For that purpose, before and after the braiding flux cycle we increase $\Phi_{0}$ from $0$ to $\Phi_{\rm max}$,  so that the spectrum of the transmon depends on the fermion parity ${\cal P}=i\gamma_A\gamma_B$ \cite{hassler2011}. During the measurement we set $\Phi_1=\Phi_2=\Phi_3=0$, to decouple the four Majoranas $\gamma_C, \gamma_D, \gamma_E, \gamma_F$ from $\gamma_A, \gamma_B$ and to minimize the effect of cross-capacitances \cite{hassler2012}.

\begin{figure*}[tb]
\centerline{\includegraphics[width=\textwidth]{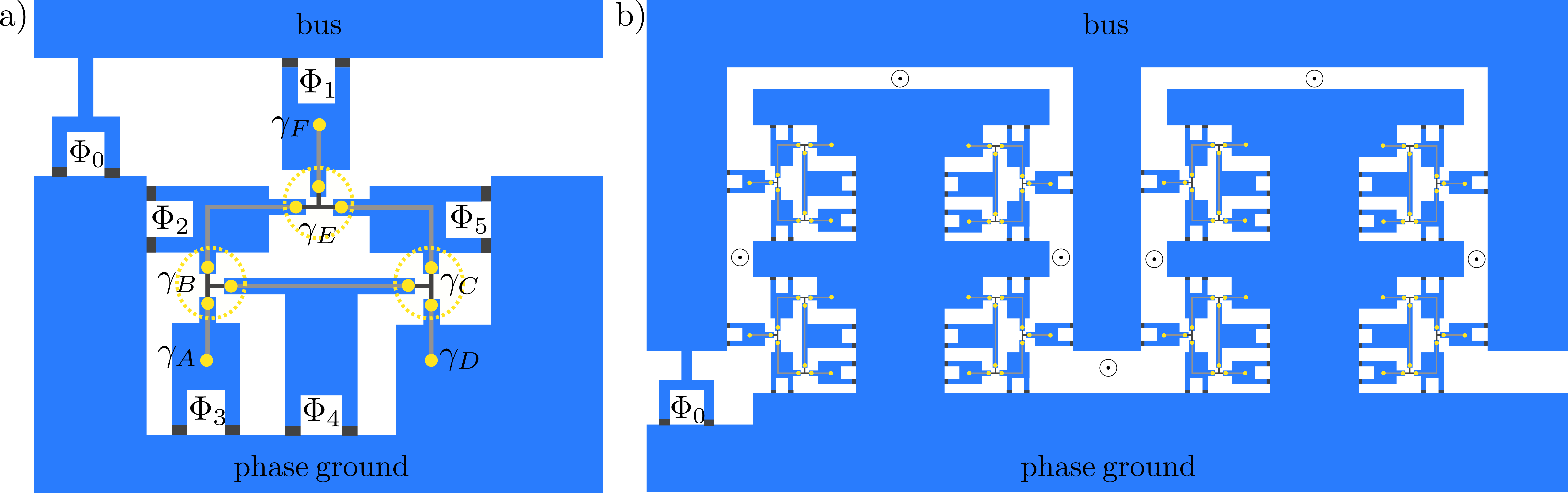}}
\caption{Panel a): Minimal transmon circuit for fully flux-controlled topological qubit. The nanowires are placed in a triangular loop formed out of three T-junctions \cite{sau2011}. In this geometry, all single-qubit Clifford gates can be implemented. Panel b): Schematic overview of a Random Access Majorana Memory consisting of eight topological qubits. Compensating fluxes (dotted circles)
are included between the topological qubits to ensure that the gauge-invariant phase differences in the different topological qubits are independent of each other (see Appendix~\ref{app_ramm}). \label{fig_layout_RAMM}}
\end{figure*}

In this configuration it is possible to execute a projective measurement on the fermion parity ${\cal P}$ by irradiating the resonator with microwaves. The system composed by the transmon qubit and microwave resonator can be described by the Hamiltonian
 \begin{align}\nonumber
H_{\text{readout}}=&\sigma_{z}\bigl[\tfrac{1}{2}\hbar\Omega_{0}+ {\cal P}\Delta_+\cos\big(\frac{\pi q_{0}}{e}\big)\bigr]+{\cal P}\Delta_{-}\cos\big(\frac{\pi q_{0}}{e}\big)\\&+\hbar \omega_{0} a^\dagger a + \hbar g(\sigma_+ a + \sigma_- a^\dagger).\label{Hreadout}
\end{align}
Here, $\omega_0$ is the bare resonance frequency of the cavity, $g$ is the strength of the coupling between photons  and the transmon qubit, and $\hbar\Omega_{0}\simeq\sqrt{8 E_{{\rm J},0}E_{{\rm C}}}$ is the transmon plasma frequency, with $E_{\rm C}$ the charging energy of the transmon including the contributions of the small islands. We have defined $\sigma_{\pm}=(\sigma_x\pm i\sigma_y)/2$ and 
\begin{equation}
\Delta_{\pm}=\frac{\delta\varepsilon_{1}\pm\delta\varepsilon_{0}}{2}  \,\frac{1}{\sqrt{1+2\cos^2(e\Phi_0/2\hbar)}} , \nonumber
\end{equation}
where $\delta\varepsilon_{1}$, $\delta\varepsilon_{0}\propto \exp(-\sqrt{8E_{{\rm J},0}/E_{C}})$ are determined by the energy levels $\varepsilon_{n}=\bar{\varepsilon}_{n}-(-1)^{n}{\delta\varepsilon}_{n} \cos(\pi q_{0}/e)$ of the transmon \cite{koch2007}. 
We assume that the induced charge is fixed at $q_0=0$ for maximal sensitivity. 

The transmission line resonator is typically operated far from resonance, in the so-called dispersive regime \cite{koch2007,dicarlo2009,houck2009}, when $(n+1)g^2\ll \delta\omega^2$, with $n$ the number of photons in the cavity and $\delta\omega=\Omega_0-\omega_0$. The Hamiltonian \eqref{Hreadout} then produces a parity-dependent resonance frequency (see Appendix \ref{app_JC})
\begin{equation}
\omega_{\rm eff}({\cal P})={}\omega_0+\sigma_z\,g^2(\delta\omega+2{\cal P} \Delta_+/\hbar)^{-1}.\label{omegaeff}
\end{equation}
 A flip of the topological qubit can thus be measured as a shift in the resonance frequency by the amount
 \begin{equation}
  \omega_{\rm shift}=\frac{4\,\hbar g^2\,\Delta_+}{\hbar^2 \delta\omega^2 - 4 \Delta_+^2}.
 \end{equation}
The probability of observing a change in the resonance frequency of the cavity after $n$ consecutive braidings, $p_{\rm flip}(n)$, is dictated by the Majorana statistics: $p_{\rm flip}(n)=\abs{\bra{1}\,{\cal U}^n\,\ket{0}}^2=\abs{\bra{0}\,{\cal U}^n\,\ket{1}}^2$. The sequence of probabilities, $p_{\rm flip}=\tfrac{1}{2}, 1, \tfrac{1}{2}, 0$ for $n=1,2,3,4$, repeats itself periodically. Therefore, the non-Abelian nature of Majoranas can be probed by collecting statistics for different values of $n$.

\section{Random Access Majorana Memory}\label{sec_RAMM}

The $\pi$-circuit of Fig.\ \ref{fig_protocol} is the minimal circuit which can demonstrate non-Abelian Majorana statistics, but it does not allow for the application of two independent braidings. The full computational power of Majoranas can be achieved by increasing the number of T-junctions. We adopt the triangular loop geometry introduced by Sau, Clarke, and Tewari \cite{sau2011}, which is the minimal circuit for a fully flux-controlled topological qubit (see Fig.~\ref{fig_layout_RAMM}a). It consists of five Majorana islands  placed between the upper and lower superconducting plates of a transmon qubit, referred to as bus and (phase) ground respectively, and a transmission line resonator for the readout.

\begin{figure*}[tb]
\centerline{\includegraphics[width=\textwidth]{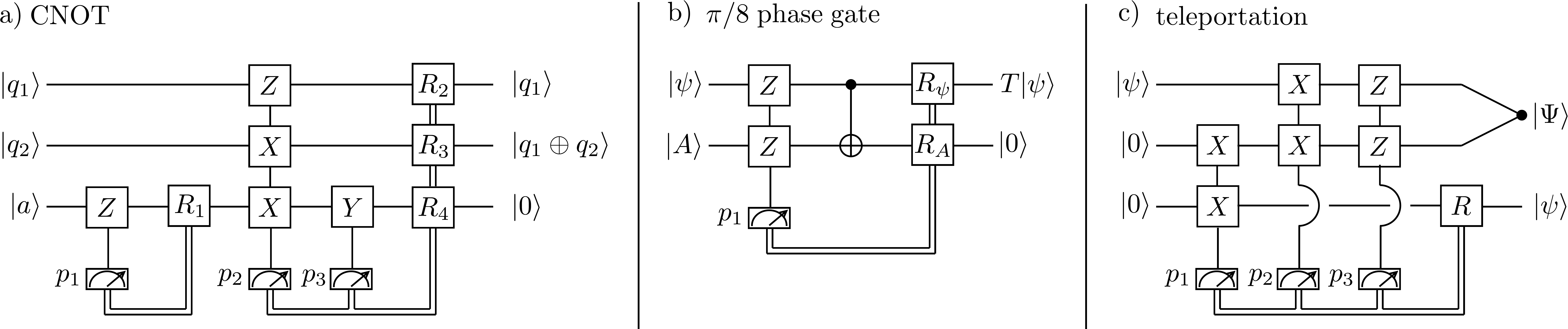}}
\caption{Quantum circuits for universal quantum computation in the {\sc ramm}. In this figure, $p_1, p_2, p_3=\pm 1$ represent results of projective single- or multi-qubit measurements, whose outcomes, carried by classical channels (double lines), determine post-selected unitary operations. Panel a): {\sc cnot} gate. Here $R_1=\exp\left[i\tfrac{\pi}{4}\sigma_x(1-p_1)\right]$, $R_2=\exp\left[i\tfrac{\pi}{4}p_2p_3\sigma_z\right]$, $R_3=\exp\left[i\tfrac{\pi}{4}p_2p_3\sigma_x\right]$, $R_4=\exp\left[-i\tfrac{\pi}{4}p_3\sigma_x\right]$ are all gates obtainable by braidings. Panel b): $\pi/8$ phase-gate $T=\diag \left( 1, \;\exp{i\tfrac{\pi}{4}}\right)$, relying on distillation of the state $\ket{A}=\left(\ket{0} + \exp i\tfrac{\pi}{4} \ket{1}\right)/\sqrt{2}$. The required unitary operations are in this case $R_\psi=\exp\left[-i\tfrac{\pi}{8}\sigma_z(1-p_
1)\right]$ and $R_A=R_1$. Panel c): teleportation protocol. Here $R=\exp\left[i\tfrac{\pi}{4}\sigma_z(1-p_1p_2)\right]\exp\left[i\tfrac{\pi}{4}\sigma_x(1-p_3)\right]$. Apart from teleporting the unknown quantum state $\ket{\psi}$, the protocol leaves the remaining two qubits in an entangled Bell state $\ket{\Psi}$. \label{gates}}
\end{figure*}

In this geometry the braiding and readout can be performed in a similar way as in the case of the $\pi$-circuit. In the braiding configuration, we set $\Phi_0=0$. Any pair of the Majoranas $\gamma_A, \gamma_B, \gamma_C$ can now be braided with the help of magnetic fluxes $\Phi_k$ ($k=1,2,...,5$). The qubit manipulations and corresponding quantum gates are shown in Appendix \ref{app_gates}. The fourth Majorana $\gamma_D$ forming the topological qubit need not be moved and is situated on the ground island, while $\gamma_E$ and $\gamma_F$ serve as ancillas. Moreover, the parity of any pair of Majoranas $\gamma_A, \gamma_B, \gamma_C$ can be measured by moving them to the ``measurement'' island, the one coupled to the bus via the flux $\Phi_1$ in Fig.~\ref{fig_layout_RAMM}a. During the measurement $\Phi_k=0$ ($k=1,2,...,5$) and $\Phi_0=\Phi_{\rm max}$, so that all the small islands are coupled via large Josephson energy either to the bus or to the ground. Therefore, the measurement configuration is described by the readout Hamiltonian \eqref{Hreadout}, where $\mathcal{P}$ is the parity of the two Majoranas in the measurement island.

Since the typical length of a transmon is hundreds of microns, it is in principle possible to scale up the design by considering a register of several topological qubits, shown in Fig.~\ref{fig_layout_RAMM}b. The measurement configuration is still described by the readout Hamiltonian \eqref{Hreadout} (see Appendix~\ref{app_ramm}), where the parity operator is now
\begin{equation}
{\cal P}=i^N\prod_{n=1}^N \gamma_{nX}\gamma_{nY}. \label{eq:multi_parity}
\end{equation}
Here $\gamma_{nX}$ and $\gamma_{nY}$ denote Majorana fermions on the measurement island belonging to topological qubit $n$: $X,Y\,\in\,\{A,B,C\}$. Thus, a readout of the resonance frequency corresponds to a projective measurement of this multi-qubit operator. Although the product in Eq.~\eqref{eq:multi_parity} runs over all $N$ qubits, we can still choose not to measure a qubit by moving the corresponding pair of coupled ancillas $\gamma_{nE}, \gamma_{nF}$ to the measurement island. Because these ancillas are always in a state $|0\rangle$, they do not influence the measurement outcome. Since the Majorana fermions can be selectively addressed, we call this architecture a Random Access Majorana Memory ({\sc ramm}). 

The number of qubits in a {\sc ramm} register cannot be increased without limitations. Firstly, the  frequency shift $\omega_{\rm shift}$ decreases with the number of topological qubits. The main decrease is caused by the reduction of the coupling $\Delta_+$ with the number of topological qubits, which occurs because the Majorana fermions at the T-junctions are localized in three different islands (see Appendix~\ref{app_ramm}). An additional decrease is caused by the renormalization of the total capacitance of the transmon due to the small islands. Furthermore, each topological qubit introduces an extra pathway for quasiparticles to be exchanged between the bus and the ground. Such quasiparticle poisoning rates at thermal equilibrium are negligibly small and the poisoning due to non-equilibrium quasiparticles can, at least in principle, be controlled by creating quasiparticle traps.

The limited number of qubits is not an obstacle for the scalability of quantum computation. Beyond this limit, the computation can be scaled up by using several transmons in a single transmission line resonator, and the coupling between the topological qubits in different registers can be achieved by introducing tunable Josephson junctions between the transmons. Furthermore, the computation can be parallelized, because transmons can be coupled to several different transmission line resonators \cite{helmer2009, divincenzo2009, mariantoni2011}.

\section{Multi-qubit measurements as a source of computational power}\label{sec_multiqubits}

Multi-qubit measurements in the {\sc ramm} offer two significant benefits. Firstly, these measurements can be applied without any locality constraint, so that the quantum fan-out \cite{divincenzo2009}, the number of other qubits with which a given qubit can interact, can become large for the {\sc ramm} architecture. Secondly,  the overhead in the computational resources can be reduced because the products of Pauli matrices involving several topological qubits can be measured directly. We demonstrate these advantages in the realization of a universal set of gates, fast creation of maximally entangled states, and implementation of error correction schemes.

\textbf{Quantum gates.} All single-qubit Clifford gates, the {\sc cnot} gate, and the $\pi/8$ phase gate required for universal quantum computation \cite{nielsen2010}, can be realized in the {\sc ramm} with errors that are exponentially small in macroscopic control parameters  (see Appendices~\ref{app_JC} and \ref{app_gates}). Single-qubit Clifford gates can be realized with braiding operations only, and the quantum circuits for the two remaining gates are summarized in Fig.~\ref{gates}. The {\sc cnot} gate, shown in Fig.~\ref{gates}a, is a modified version of the Bravyi-Kitaev algorithm \cite{bravyi2002,bravyi2006} involving three topological qubits (target, control, and one ancilla). Efficient $\pi/8$ phase gate implementations are based on distillation protocols \cite{bravyi2005}, requiring several noisy qubits to prepare one qubit in a particular state $|A\rangle=\left( \ket{0} + \e^{i\pi/4} \ket{1}\right)/\sqrt{2}$. This state can then be used to perform the $\pi/8$ gate using the circuit shown in Fig.~\ref{gates}
b. Distillation may take place in  dedicated {\sc ramm} registers (see Appendix~\ref{app_gates}) in parallel with other computation processes, and the distilled state can be teleported to the computational register (see Fig.~\ref{gates}c). 

\begin{figure}
\includegraphics[width=\columnwidth]{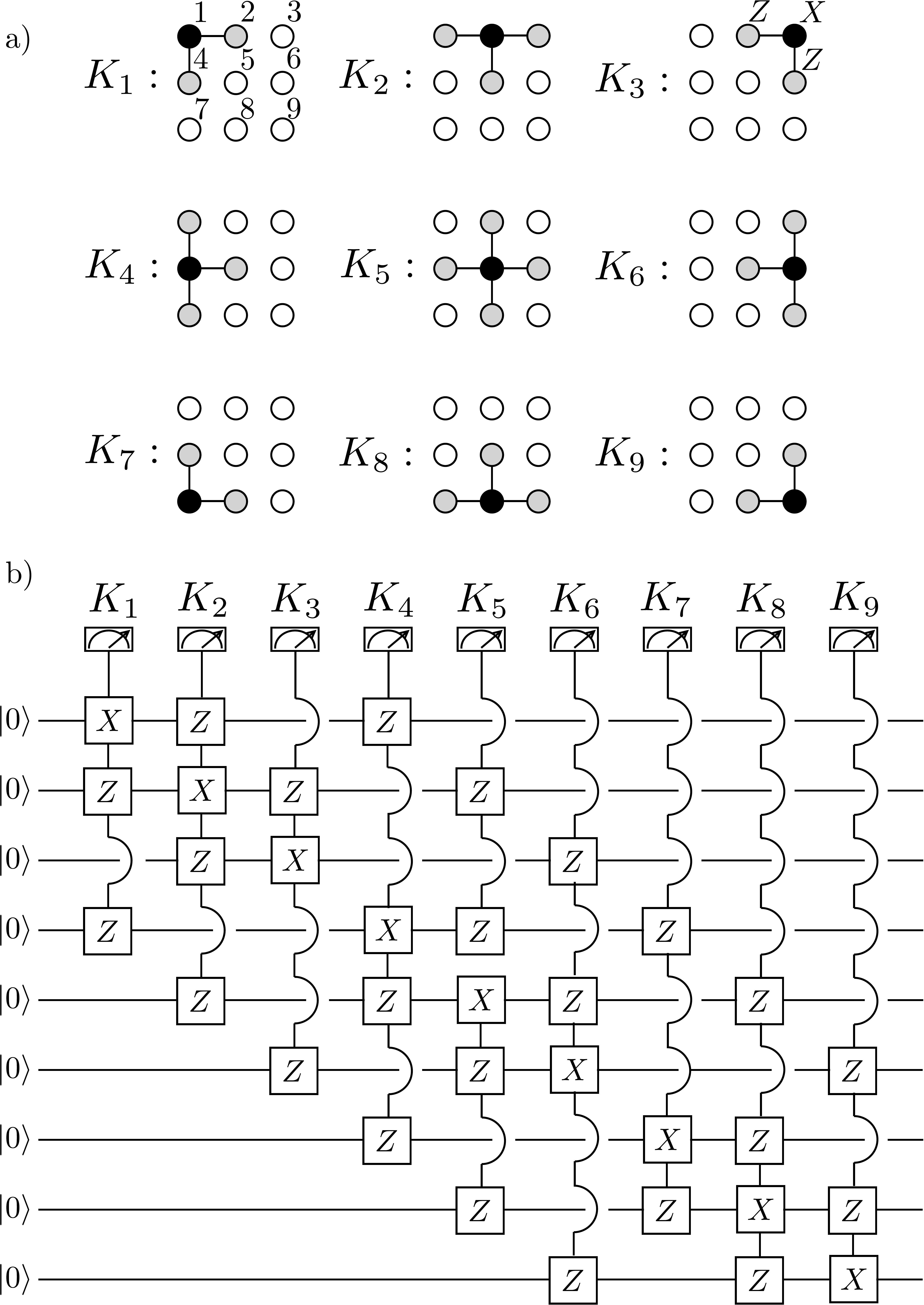} 
\caption{Preparation of a 9-qubit 2D cluster state with a {\sc ramm}. The nine qubits (represented by circles) are arranged in a $3\times 3$ square logical lattice, and numbered from left to right and top to bottom. Panel a): The nine stabilizer operators $K_1,\dots K_9$ necessary to prepare the 2D cluster state. They are products of Pauli matrices, involving all qubits connected by
lines, with black and grey dots representing $\sigma_x$ and $\sigma_z$ operators, respectively. Panel b): The quantum circuit creating the 2D cluster state in a 9-qubit {\sc ramm} register, consisting in a sequence of projective multi-qubit measurement of the 9 stabilizers. \label{2dcluster}}
\end{figure}

\textbf{Preparation of 2D cluster states.} The {\sc ramm} can be used to efficiently create maximally entangled multi-qubit states, such as 2D cluster states \cite{hein2006,briegel2009,briegel2001}, which make it possible to realize any quantum circuit by means of single-qubit operations and measurements \cite{raussendorf2001}.

To generate a 2D cluster state in the {\sc ramm} architecture one has first to assign a label to each topological qubit in order to establish its position and neighbours on a \textit{logical lattice} (see Fig.\ \ref{2dcluster}a). Due to the non-locality of measurements in the {\sc ramm}, the logical lattice does not need to be related to the physical system. The cluster state may be prepared in several ways \cite{hein2006,briegel2001}. An efficient procedure requires measuring the stabilizers 
\begin{equation}
 K_\alpha = \sigma_{x,\alpha} \prod_{\langle \beta, \alpha \rangle} \sigma_{z,\beta}\label{eq:stabilizers},
\end{equation}
where $\alpha$ goes through all sites of the logical lattice and $\beta$ labels the nearest neighbours of $\alpha$. The total number of measurements required is equal to the number of qubits in the cluster state. In Fig.~\ref{2dcluster}b we draw a circuit to create the 9-qubit 2D cluster state in a {\sc ramm} register. To verify their entanglement properties, one possibility is provided by the teleportation protocol of Ref.~\onlinecite{raussendorf2001}.

\textbf{Efficient quantum error correction.} Although topological qubits have intrinsically low error rates, grouping them into a {\sc ramm} register allows to additionally implement efficient error correction. Error correction schemes \cite{preskill1998,nielsen2010} are based on measurements of stabilizer generators, which are products of Pauli matrices belonging to different qubits. The measurement outcomes give error syndromes, which uniquely characterize the errors and the qubits where they occurred. The {\sc ramm} allows for efficient error correction schemes, due to the possibility of measuring stabilizers of different length, as well as correcting errors using single-qubit Clifford gates. There are two advantages in comparison with architectures where only single- and two-qubit operations are available: higher error thresholds and reduced overhead in computational resources.

In order to quantitatively compare these advantages, we consider the 7-qubit Steane code \cite{steane1996} as a concrete example of quantum codes, and assume a realistic error model. 
We find that the error threshold of the {\sc ramm} can be  an order of magnitude larger than the error threshold of a reference architecture that can only perform single- and two-qubit operations (see Appendix~\ref{app_thresholds}).
Additionally, the {\sc ramm} implementation of the Steane code is much more compact. Already in the first level of concatenation, the fault-tolerant implementation of syndrome measurements in the reference architecture requires 24 ancillas for each logical qubit, while none are needed in the {\sc ramm}.

Although we have calculated the improvements only for the 7-qubit Steane code, the advantages are characteristic for all error correction schemes, including surface codes \cite{bravyi1998,fowler2012}.

\section{Discussion}\label{sec_discussion}

To control and manipulate quantum information contained in the Majorana zero-modes of superconducting nanowires it is necessary to braid them and measure their parity. 
We have designed a transmon circuit where both operations can be performed by controlling the magnetic fluxes through split Josephson junctions, without local adjustment of microscopic parameters of the nanowires. The minimal circuit for the demonstration of non-Abelian Majorana statistics is a $\pi$-shaped circuit involving four independent flux variables. An extended circuit consisting of many topological qubits in parallel allows for non-local multi-qubit measurements in a Random Access Majorana Memory, providing the possibilities of efficient creation of highly entangled states and simplified (ancilla-free) quantum error correction.

Since all the requirements for the realization of the $\pi$-circuit and  {\sc ramm} are satisfied with the typical energy scales of existing transmon circuits  and  transmission line resonators (see Appendix~\ref{app_energies}), flux-controlled circuits are a favorable architecture for the demonstration of non-Abelian Majorana statistics and the realization of fault-tolerant quantum computation.

\acknowledgments

We have benefited from discussions with E. Alba. This work was supported by the Dutch Science Foundation NWO/FOM, by an ERC Advanced Investigator Grant, and by a Lawrence Golub Fellowship.

\appendix

\section{Theoretical description of the $\pi$-shaped circuit}\label{app_pi_circ}

\begin{figure}[t]
\centerline{\includegraphics[width=\linewidth]{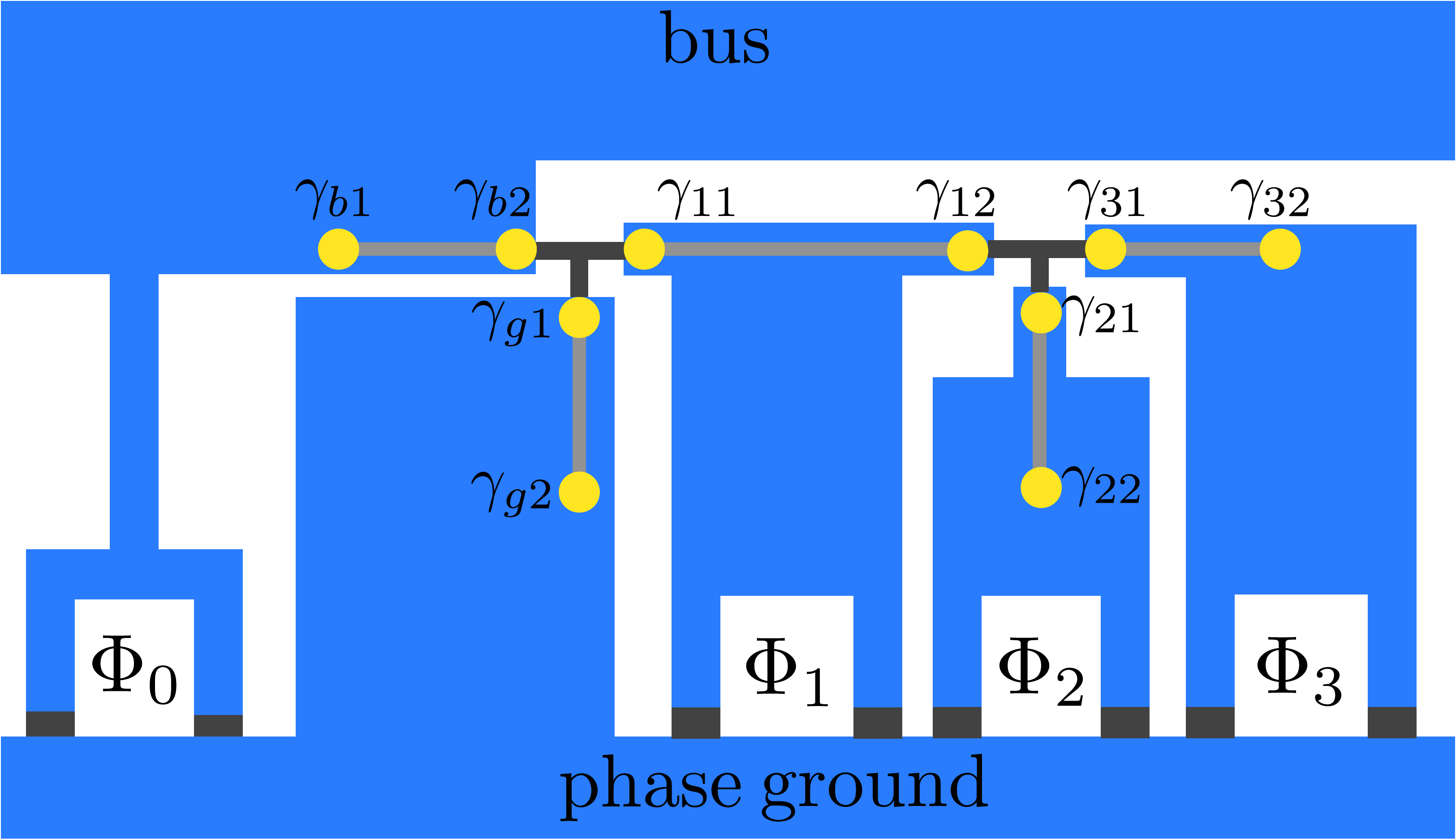}}
\caption{The $\pi$-shaped transmon circuit discussed in the main text, reproduced here with labels of the ten Majorana bound states.}
\label{fig:pi_circuit}
\end{figure}

The $\pi$-shaped circuit discussed in the main text is reproduced here in Fig.\ \ref{fig:pi_circuit}. We label the two superconducting plates forming the transmon ``bus'' and ``ground'', both hosting two Majorana bound states, labeled $\gamma_{b1}, \gamma_{b2}$ and $\gamma_{g1},\gamma_{g2}$ respectively. The smaller superconducting islands are labeled with an integer $k=1,2,3$. Each of them supports two Majorana bound states $\gamma_{k1}, \gamma_{k2}$. We will work in a gauge where all phases are measured with respect to the phase of the ground island. We denote with $\phi$ the phase of the bus and with $\phi_k$ that of the $k$-th island.

We start from the Lagrangian of the system,
\begin{equation}\label{lagrangian}
{\cal L}=T-V_J-V_M\,.
\end{equation}
The first term is the charging energy
\begin{align}\notag
T&=\frac{\hbar^2}{8e^2}C_0\dot\phi^2 + \frac{\hbar^2}{8e^2} \sum_{k=1}^3 \left[C_{G,k}\, \dot\phi_k^2 + C_{B,k}\left(\dot\phi_k-\dot\phi\right)^2\right]\\&+ \frac{\hbar}{2e} \left[q_0\dot\phi+\sum_{k=1}^3 q_k\dot\phi_k\right]. \label{kinetic}
\end{align}
Here $C_0$ is the capacitance between bus and ground, while $C_{G,k}$ ($C_{B,k}$) is the capacitance between the $k$-th Majorana island and the ground (the bus). The last two terms include the induced charge $q_0$  on the bus and $q_k$ on Majorana islands. The effect of cross-capacitances between Majorana islands is negligible assuming that they are small in comparison with the capacitances to the bus and the ground. 

The second term is the Josephson potential 
\begin{equation}\notag
V_J=E_{J,0}(\Phi)\,(1-\cos\phi)+\sum_{k=1}^3\,E_{J,k}(\Phi_k)(1-\cos\phi_k).
\end{equation}
The Josephson energies $E_{J,0}(\Phi_0)=2E_{J,0}(0)\cos(e\Phi_0/\hbar)$ and $E_{J,k}(\Phi_k)=2E_{J,k}(0)\cos(e\Phi_k/\hbar)$ can be varied in magnitude by changing the fluxes between $0$ and $\abs{\Phi_{\rm max}}\lesssim h/4e$. We are assuming for simplicity that the split junctions are symmetrical, but this requirement can be removed without affecting our results.

The third term is the Majorana-Josephson potential
\begin{align}\label{VM}
V_M=E_M&\left[i\gamma_{b2}\gamma_{g1}\cos\left(\tfrac{1}{2}\phi+\alpha_{bg}\right)\right.\\\nonumber
			&+i\gamma_{g1}\gamma_{11}\cos\left(\alpha_{g1}-\tfrac{1}{2}\phi_1\right)\\\nonumber
			&+\left.i\gamma_{11}\gamma_{b2}\cos\left(\tfrac{1}{2}\phi_1-\tfrac{1}{2}\phi+\alpha_{1b}\right)\right]\\\nonumber
			+E_M&\left[i\gamma_{12}\gamma_{21}\cos\left(\tfrac{1}{2}\phi_1-\tfrac{1}{2}\phi_2+\alpha_{12}\right)\right.\\\nonumber
			&+i\gamma_{21}\gamma_{31}\cos\left(\tfrac{1}{2}\phi_2-\tfrac{1}{2}\phi_3+\alpha_{23}\right)\\\nonumber
			&+\left.i\gamma_{31}\gamma_{12}\cos\left(\tfrac{1}{2}\phi_3-\tfrac{1}{2}\phi_1+\alpha_{31}\right)\right].
\end{align}
The two square brackets in this expression group the terms corresponding to the two T-junctions. All tunnel couplings are for simplicity assumed to be of equal strength $E_M$. The arguments of the cosines include single-electron Aharonov-Bohm phase shifts between different islands, 
\begin{subequations}
\begin{align}
&\alpha_{bg}=e\Phi_0/2\hbar\\
&\alpha_{g1}=e\Phi_1/2\hbar\\
&\alpha_{1b}=-\left(e\Phi_0+e\Phi_1\right)/2\hbar\\
&\alpha_{12}=(e\Phi_1+e\Phi_2)/2\hbar\\
&\alpha_{23}=(e\Phi_2+e\Phi_3)/2\hbar\\
&\alpha_{31}=-\left(e\Phi_1+2e\Phi_2+e\Phi_3\right)/2\hbar
\end{align}
\end{subequations}

There is a constraint between the charge contained in each superconducting island and the parity of the Majorana fermions belonging to that island \cite{fu2010}. The constraint can be eliminated via a gauge transformation \cite{heck2011}
\begin{align}
\Omega&=\e^{in\phi/2}\prod_{k=1}^3\e^{in_{k}\phi_{k}/2}\label{gaugetransformation}\\
n&=\tfrac{1}{2}-\tfrac{1}{2}i\gamma_{b1}\gamma_{b2}\;,\;\;n_k=\tfrac{1}{2}-\tfrac{1}{2}i\gamma_{k1}\gamma_{k2}\,,
\end{align}
where the products extends over all Majorana junctions. The transformation has two effects on the Lagrangian:
\begin{itemize}
	\item it changes the induced charges appearing in Eq. \eqref{kinetic},
	\begin{equation}\label{inducedcharges}
	q_0 \to q_0+e n\;,\;q_k\to q_k+en_k
	\end{equation}
	so that the Majorana operators enter explicitly in the charging energy, and
	\item it modifies the Majorana-Josephson potential $\Omega^\dagger V_M \Omega$ so that it becomes $2\pi$-periodic in all its arguments $\phi, \phi_k$.
\end{itemize}
In the following, we will work in this new gauge where Eq. \eqref{inducedcharges} holds. The explicit form of $\Omega^\dagger V_M \Omega$ is not necessary here, as we will only need the equality
\begin{equation}
\left.\Omega^\dagger V_M \Omega\right|_{\phi_k=\phi=0}=\left.V_M\right|_{\phi_k=\phi=0}
\end{equation}
which is trivial since $\left.\Omega\right|_{\phi_k=\phi=0}=1$. Starting from the Lagrangian \eqref{lagrangian}, we will now derive the low-energy Hamiltonians used in the main text for the braiding and the readout.

\subsection{Braiding} 

When we want to braid or move the Majoranas, we maximize the energy $E_{J,0}(\Phi_0)$ by setting $\Phi_0=0$ and we require the condition
\begin{equation}
E_{J,0}(0), E_{J,k}(\Phi_k) \gg E_M, E_C, E_{C,k}
\end{equation}
where $E_{C,0}=e^2/2C_0$ and $E_{C,k}=e^2/2({C_{B,k}+C_{G,k}})$. Since the Josephson term $V_J$ dominates over the kinetic and Majorana terms $T$ and $V_M$, the action $S=\int\mathcal{L}\de t$ is then minimized for $\phi=\phi_k=0$ and $\dot\phi=\dot\phi_k=0$. All the superconducting islands are in phase. Under the additional condition
\begin{equation}
\frac{E_{J,0}(0)}{E_{C,0}}>\frac{E_{J,k}(\Phi_{k})}{E_{C,k}}\;,
\end{equation}
we can neglect quantum phase slips around the minimum $\phi=0$, but not around the other minima $\phi_k=0$. The low-energy Hamiltonian $H_{\rm M}$ then contains only the Majorana operators:
\begin{equation}
H_{\rm eff}=-\sum_{k=1}^{3}iU_{k}\gamma_{k1}\gamma_{k2}+\left.\Omega^\dagger V_M\Omega\right|_{\phi_k=\phi=0}\label{effective-Ham}
\end{equation}
where
\begin{equation}
U_k = 16\left(\frac{E_{{\rm C},k} E_{{\rm J},k}^3}{2\pi^2}\right)^{\tfrac{1}{4}}\e^{-\sqrt{8E_{{\rm J},k}/E_{{\rm C},k}}} \cos(q_k\pi/e), \label{Uksupp}
\end{equation}
is the tunneling amplitude of a phase slip process from $\phi_k=0$ to $\phi_k=\pm 2\pi$ \cite{koch2007}, also reported in Eq. (1) of the main text. 

There are still ten Majorana operators in the Hamiltonian \eqref{effective-Ham}, but we can eliminate four of them by assuming that the tunnel couplings are stronger than the Coulomb couplings: $E_{\rm M}\gg U_{k}$. To first order in perturbation theory in the ratio $U_k/E_{\rm M}$, we then obtain the Hamiltonian used in the main text
\begin{equation}
H=-i\Delta_{1}\gamma_{B}\gamma_{E}-i\Delta_{2}\gamma_{E}\gamma_{F}-i\Delta_{3}\gamma_{E}\gamma_{C}
\end{equation}
In this passage we have introduced the six Majorana operators $\gamma_A, \gamma_B,\gamma_C, \gamma_D, \gamma_E, \gamma_F$, given by
\begin{subequations}
\label{gammasix}
\begin{align}
&\gamma_A=\gamma_{b1},\\
&\gamma_B=\frac{\cos\alpha_{g1}\gamma_{b2}+\cos\alpha_{1b}\gamma_{g1}+\cos\alpha_{bg}\gamma_{11}}{\sqrt{\cos^2\alpha_{g1}+\cos^2\alpha_{1b}+\cos^2\alpha_{bg}}},\\
&\gamma_C=\gamma_{32},\\
&\gamma_D=\gamma_{g2},\\
&\gamma_E=\frac{\cos\alpha_{23}\gamma_{12}+\cos\alpha_{31}\gamma_{21}+\cos\alpha_{12}\gamma_{31}}{\sqrt{\cos^2\alpha_{23}+\cos^2\alpha_{31}+\cos^2\alpha_{12}}},\\
&\gamma_F=\gamma_{22}.
\end{align}
\end{subequations}
The coupling strengths are
\begin{subequations}
\begin{align}
\Delta_1 ={}& U_1\frac{\cos\alpha_{bg}}{\sqrt{\cos^2\alpha_{g1}+\cos^2\alpha_{1b}+\cos^2\alpha_{bg}}}  \nonumber \\
  & \times \frac{\cos\,\alpha_{23}}{\sqrt{\cos^2\alpha_{12}+\cos^2\alpha_{23}+\cos^2\alpha_{31}}},\\
\Delta_2 ={}& U_2\frac{\cos\,\alpha_{31}}{\sqrt{\cos^2\alpha_{12}+\cos^2\alpha_{23}+\cos^2\alpha_{31}}},\\
\Delta_3 ={}& U_3\frac{\cos\,\alpha_{12}}{\sqrt{\cos^2\alpha_{12}+\cos^2\alpha_{23}+\cos^2\alpha_{31}}}.
\end{align}
\end{subequations}

\subsection{Readout}\label{readout}

During the readout of the transmon qubit, we set $\Phi_0=\Phi_{\rm max}$, so that the Josephson energy $E_{J,0}$ is minimized, and all $\Phi_k=0$. We require then that
\begin{equation}
\frac{E_{J,k}(0)}{E_{C,k}}\gg \frac{E_{J,0}(\Phi_{\rm max})}{E_{C,0}}.
\end{equation}
In physical terms, all Majorana islands are now in phase with the ground: $\phi_k=\dot\phi_k=0$. Neglecting quantum fluctuations and phase slips around these minima, we may re-write the Lagrangian in a form that depends only on $\phi$
\begin{align}\notag
\mathcal{L}=&\frac{\hbar^2}{8e^2}C \dot\phi^2+\frac{\hbar}{2e}(q_0+en)\dot\phi\\
&-E_{J, 0}\,(1-\cos \phi)-\left.\Omega^\dagger V_M\Omega \right|_{\phi_k=0}
\end{align}
Apart from the contribution of the term $V_M$, the whole system can be treated as a single hybrid top-transmon \cite{hassler2011}, with Josephson energy $E_{J,0}$ and capacitance 
\begin{align}
C&=C_0+\sum_{k=1}^3 C_{B,k}.
\end{align}
In the regime $E_{J,0} \gg E_C=e^2/2C$, the energy levels of the transmon are given by \cite{koch2007}
\begin{equation}
\varepsilon_{n}=\bar{\varepsilon}_{n}-(-1)^{n}{\delta\varepsilon}_{n} i\gamma_{b1}\gamma_{b2} \cos(\pi q/e)\;,
\end{equation}
where
\begin{align}
&\bar{\varepsilon}_{n}\simeq -E_{J,0}+\left(n+\tfrac{1}{2}\right) \sqrt{8E_{J,0} E_C}-\frac{E_C}{12}(6n^2+6n+3) \label{anharmonic} \\
&\delta\varepsilon_{n}=E_C\frac{2^{4n+4}}{n!}\sqrt{\frac{2}{\pi}}\left(\frac{E_{J,0}}{2E_C}\right)^{\tfrac{n}{2}+\tfrac{3}{4}}\e^{-\sqrt{8E_{J,0}/E_C}} \label{transshift}.
\end{align}

Taking into account the two lowest levels of the transmon ($n=0,1$), we arrive at a low-energy Hamiltonian
\begin{align}
H_\textrm{top-transmon}&=\sigma_{z}\bigl[\tfrac{1}{2}\hbar\Omega_{0}+i\gamma_{b1}\gamma_{b2}\,\delta_{+}\cos(\pi q_0/e)\bigr]\nonumber\\+i\gamma_{b1}\gamma_{b2}\,&\delta_{-}\cos(\pi q_0/e)+\left.\Omega^\dagger V_M\Omega\right|_{\phi_k=\phi=0}
\end{align}
with definitions $\hbar\Omega_{0}= \bar{\varepsilon}_{1}-\bar{\varepsilon}_{0}$, $\delta_{\pm}=(\delta\varepsilon_{1}\pm\delta\varepsilon_{0})/2$. The Pauli matrix $\sigma_{z}$ acts on the qubit degree of freedom of the transmon. For $\delta_\pm\ll E_M$,  the low energy sector of this Hamiltonian can be written in terms of $\gamma_A, \dots, \gamma_F$ as
\begin{align}\label{Htt}
\tilde{H}_\textrm{top-transmon}&=\sigma_{z}\bigl[\tfrac{1}{2}\hbar\Omega_{0}+i\gamma_{A}\gamma_{B}\,\Delta_{+}\cos(\pi q_0/e)\bigr]\nonumber\\&\qquad+i\gamma_{A}\gamma_{B}\,\Delta_{-}\cos(\pi q_0/e)
\end{align}
where
\begin{equation}
\Delta_{\pm}=\frac{\delta_{\pm} \cos\alpha_{g1}}{\sqrt{\cos^2\alpha_{bg}+\cos^2\alpha_{g1}+\cos^2\alpha_{1b}}}.
\end{equation}
When combined with the Jaynes-Cummings Hamiltonian describing the coupling with the resonator, this Hamiltonian reproduces Eq. (5) of the main text. The interaction with the microwaves will be described in detail in the next Appendix \ref{app_JC}.

\section{Measurement through photon transmission}\label{app_JC}

The Hamiltonian $H_{\text{readout}}$ of the main text describes the coupling between the top-transmon and the cavity modes in the system through a Jaynes-Cummings interaction of strenght $g$. In particular the fermionic parity of the transmon $\mathcal{P}$ is a conserved quantity in the Hamiltonian whose energy levels will directly depend on the value of $\mathcal{P}$.

We assume that the induced charge is fixed at $q_0=0$ to maximize the sensitivity of the read-out. The Jaynes-Cummings interaction couples the pairs of states $\left( \ket{n,\uparrow,\mathcal{P}},\ket{n+1,\downarrow,\mathcal{P}}\right) $ where $n$ and $n+1$ label the number of photons in the cavity and $\ket{\uparrow},\ket{\downarrow}$ denote the two lowest energy eigenstates of the transmon. Therefore, the eigenstates of $H_{\text{readout}}$ are in general superpositions of the kind $\alpha \ket{n,\uparrow,\mathcal{P}}+\beta \ket{n+1,\downarrow,\mathcal{P}}$ with the exception of the uncoupled vacuum states $\ket{0,\downarrow,\mathcal{P}}$. Their eigenvalues are, respectively:
\begin{align}
&\varepsilon_{n,\pm,\mathcal{P}} = \left(n+\frac{1}{2}\right) \hbar\omega_0 +\mathcal{P}\Delta_- +  \nonumber \\
&\qquad \qquad  \pm \frac{1}{2} \sqrt{(\hbar\delta\omega + 2\mathcal{P}\Delta_+)^2+ 4\hbar^2 g^2\left(n+1 \right)} \\
&\varepsilon_{0,\mathcal{P}} =  \mathcal{P}\left(\Delta_- - \Delta_+\right) -\frac{1}{2}\hbar\Omega_0. 
\end{align}
In the dispersive regime, $\delta\omega^2 \gg g^2(n+1)$, the energies $\varepsilon_{n,\pm,\mathcal{P}}$ can be approximated at the first order in $g^2/\delta\omega^2$ as:
\begin{align}
 &\varepsilon_{n,\uparrow,\mathcal{P}} = n\hbar \omega_0 + \mathcal{P}\left(\Delta_-+\Delta_+ \right) +\frac{1}{2} \hbar\Omega_0 +\frac{\hbar^2g^2\left( n+1\right) }{\hbar\delta\omega +2\mathcal{P}\Delta_+} \\
 &\varepsilon_{n+1,\downarrow,\mathcal{P}} = (n+1)\hbar \omega_0 + \mathcal{P}\left(\Delta_--\Delta_+ \right) + \nonumber \\
&\qquad \qquad \qquad \qquad \qquad -\frac{1}{2} \hbar\Omega_0 -\frac{\hbar^2g^2\left( n+1\right) }{\hbar\delta\omega +2\mathcal{P}\Delta_+}.
\end{align}
The respective eigenstates are approximately $\ket{n,\uparrow,\mathcal{P}}$ and $\ket{n+1,\downarrow,\mathcal{P}}$ up to corrections of the order $g^2/\delta\omega^2$. From the previous equations it is easy to obtain the effective resonance frequency $\omega_{\rm eff}\left( \mathcal{P}\right)$ and its shift $\omega_{\rm shift}$ corresponding to the different states of the topological qubit. Since we are considering the dispersive regime with a positive detuning, $\Omega_0 > \omega_0$, we assume in the following that the state of the transmon remains in the ground state $\ket{\downarrow}$. 

We also point out that in the Hamiltonian $H_{\text{readout}}$ we are neglecting the excited states of the transmon, which result in a renormalization of the parameters, including $\omega_{\rm shift}$, through virtual transitions. The precise expressions for the renormalized parameters are known \cite{koch2007}, but are not needed here.  

To perform the measurement of the topological qubit we introduce in the cavity photons with a frequency which is approximately $\omega_{\rm eff}(\mathcal{P}=+1)$. The photon transmission probability $T_+$ for the state $\ket{\mathcal{P}=1}$ is then larger than the probability $T_-$ corresponding to $\ket{\mathcal{P}=-1}$. We count the number of photons $n_{\textrm{ph}}$ that passes through the cavity during a measurement time $t_M$. The probability distributions for $n_{\textrm{ph}}$ in each state are Poissonian, and for sufficiently long measurement time can be approximated with normal distributions 
\begin{equation}
 {\mathbb P} (n_{\textrm{ph}}, |\mathcal{P}=\pm 1\rangle) = {\sf Pois} (n_{\textrm{ph}}, \lambda_\pm) \approx {\mathbb N} (n_{\textrm{ph}},\lambda_\pm, \sqrt{\lambda_\pm}) 
\end{equation}
where $\lambda_\pm  \propto T_\pm t_M \kappa$ and $\kappa \simeq 1-10\,\text{MHz}$ is the cavity decay rate. Since $T_+ > T_-$, also $\lambda_+ > \lambda_-$.

We decide that the measurement outcome is $\mathcal{P}=+1$ if $n_{\textrm{ph}}> x=\sqrt{\lambda_+\lambda_-}$ and the outcome is $\mathcal{P}=-1$ if $n_{\textrm{ph}}< x$. Therefore the error of the measurement outcome is given by the following:
\begin{multline}
 \epsilon_{\textrm{om}} = \frac{1}{2} \int_{-\infty}^x \frac{dn}{\sqrt{2\pi\lambda_+}}\exp\left( \frac{-(n-\lambda_+)^2}{2\lambda_+} \right) + \\
+ \frac{1}{2}  \int_x^{\infty} \frac{dn}{\sqrt{2\pi\lambda_-}}\exp\left( \frac{-(n-\lambda_-)^2}{2\lambda_-} \right).
\end{multline}
Since $\lambda_+, \ \lambda_- \gg 1$
\begin{equation}
 \epsilon_\textrm{om} \simeq \frac{e^{-{\bar x}^2}}{2{\bar x}\sqrt{\pi}},
 \end{equation}
where 
\begin{equation}
\bar{x}= \frac{\sqrt{\lambda_+} - \sqrt{\lambda_-}}{\sqrt{2}}.
\end{equation}
We notice that the probability of a measurement error decreases exponentially with $\kappa t_M$. On the other hand, the probability of \textit{storage} error, namely the chance that the topological qubit will decay during a time interval $t_M$, increases as $\Delta_\text{min} t_M/\hbar$. Because $\Delta_\text{min}/\kappa$ can be made exponentially small in macroscopic control parameters, exponentially small measurement errors can be achieved. 

\section{Low energy Hamiltonian for a Random Access Majorana Memory architecture}\label{app_ramm}

\begin{figure*}[t]
\centerline{\includegraphics[width=0.75\textwidth]{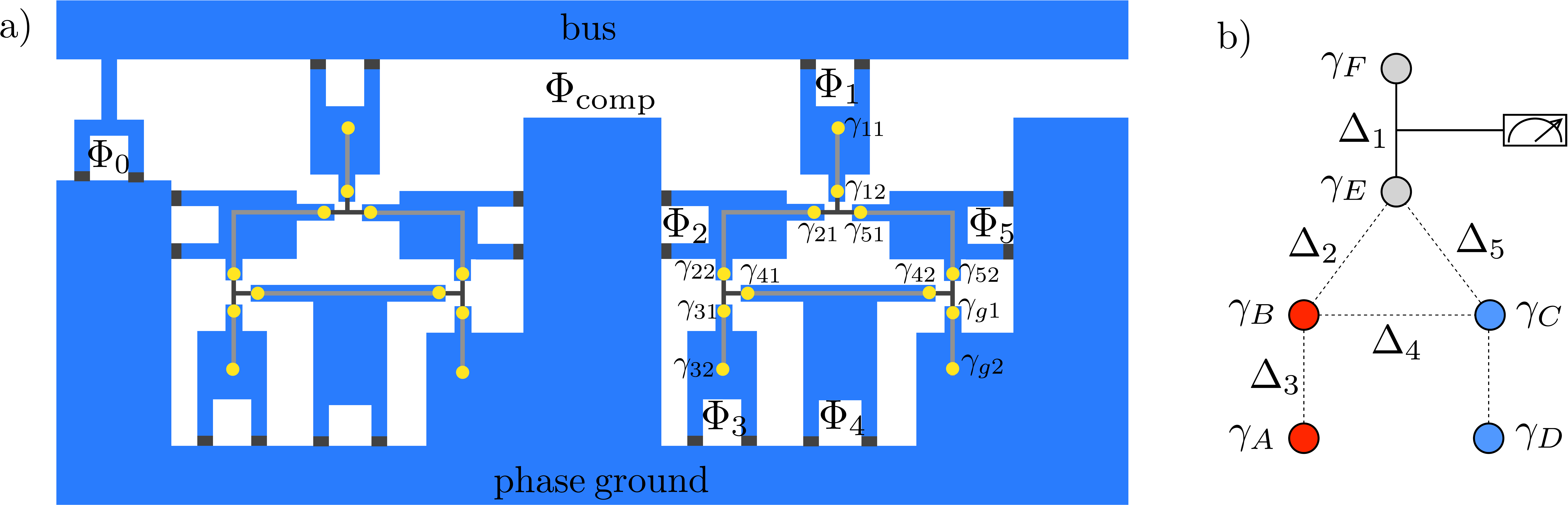}}
\caption{Panel (a): Part of the {\sc ramm} circuit showing two fully-controllable topological qubits.  Compensating fluxes are included between the topological qubits in order that the gauge-invariant phase differences in the different topological qubits are independent of each other. Panel (b): Topological qubit formed by the six Majorana fermions. The five couplings $\Delta_1, \dots, \Delta_5$, see Eq. \eqref{qubit_couplings}, can all be individually controlled by the fluxes $\Phi_1, \dots, \Phi_5$. The parity of the two Majoranas coupled by $\Delta_1$ can be measured, as explained in Appendix~\ref{RAMMreadout}.}
\label{fig:top_qubit}
\end{figure*}

We will now describe an effective Hamiltonian for {\sc ramm} architecture hosting $N$ topological qubits, such as the one shown in Fig.\ 3 of the main text. Fig.~\ref{fig:top_qubit}a shows an equivalent setup, including only two topological qubits. By including compensating fluxes
\begin{equation}
\Phi_{\textrm{comp}, n}=-\sum_{k=1}^5 \Phi_{n,k}
\end{equation}
after each topological qubit, the gauge invariant phases in each topological qubit are independent of each other. The single-electron Aharonov-Bohm phase-shifts $\alpha_{n,kk'}$ at the tunnel junction between islands $k$ and $k'$ of the $n$-th qubit are then given by
\begin{align}\nonumber
\alpha_{n,12}&=e(\Phi_{0}+\Phi_{n,1}+\Phi_{n,2})/2\hbar\\\nonumber
\alpha_{n,25}&=e(\Phi_{n,2}+2\Phi_{n,3}+2\Phi_{n,4}+\Phi_{n,5})/2\hbar\\\nonumber
\alpha_{n,51}&=-e(\Phi_{0}+\Phi_{n,1}+2\Phi_{n,2}+2\Phi_{n,3}\\\nonumber
&\qquad\qquad\qquad+2\Phi_{n,4}+\Phi_{n,5})/2\hbar\\\nonumber
\alpha_{n,23}&=e(\Phi_{n,2}+\Phi_{n,3})/2\hbar\\\nonumber
\alpha_{n,34}&=e(\Phi_{n,3}+\Phi_{n,4})/2\hbar\\\nonumber
\alpha_{n,42}&=-e(\Phi_{n,2}+2\Phi_{n,3}+\Phi_{n,4})/2\hbar\\\nonumber
\alpha_{n,4g}&=e\Phi_{n,4}/2\hbar\\\nonumber
\alpha_{n,g5}&=e\Phi_{n,5}/2\hbar\\
\alpha_{n,54}&=-e(\Phi_{n,4}+\Phi_{n,5})/2\hbar. \label{phases}
\end{align}
Here, the subscript $g$ denotes the tunnel junctions to the ground island.

By starting from a Lagrangian and following a similar approach to that of Appendix \ref{app_pi_circ}, we find
that the low-energy Hamiltonian is described by six Majorana fermions 
\begin{align}\nonumber
&\gamma_{n,A}=\gamma_{n,32},\\\nonumber
&\gamma_{n,B}=\frac{\cos\alpha_{n,34}\gamma_{n,22}+\cos\alpha_{n,42}\gamma_{n,31}+\cos\alpha_{n,23}\gamma_{n,41}}{\sqrt{\cos^2\alpha_{n,23}+\cos^2\alpha_{n,34}+\cos^2\alpha_{n,42}}},\\\nonumber
&\gamma_{n,C}=\frac{\cos\alpha_{n,g5}\gamma_{n,42}+\cos\alpha_{n,54}\gamma_{n,g1}+\cos\alpha_{n,4g}\gamma_{n,52}}{\sqrt{\cos^2\alpha_{n,4g}+\cos^2\alpha_{n,g5}+\cos^2\alpha_{n,54}}},\\\nonumber
&\gamma_{n,D}=\gamma_{n,g2},\\\nonumber
&\gamma_{n,E}=\frac{\cos\alpha_{n,25}\gamma_{n,12}+\cos\alpha_{n,51}\gamma_{n,21}+\cos\alpha_{n,12}\gamma_{n,51}}{\sqrt{\cos^2\alpha_{n,12}+\cos^2\alpha_{n,25}+\cos^2\alpha_{n,51}}},\\
&\gamma_{n,F}=\gamma_{n,11}\,.
\end{align}
that form the triangular loop network of Fig.\ \ref{fig:top_qubit}b.

\subsection{Low-energy Hamiltonian in braiding configuration}

In the braiding configuration $\Phi_0=0$, and the low-energy Hamiltonian is, for each qubit $n$,
\begin{align}\nonumber
H^{(n)}_\textrm{qubit}=&-i\Delta_{n,1}\gamma_F\gamma_E-i\Delta_{n,2}\gamma_E\gamma_B-i\Delta_{n,3}\gamma_B\gamma_A\\
				&-i\Delta_{n,4}\gamma_B\gamma_C-i\Delta_{n,5}\gamma_E\gamma_C\,,\label{Hqubit}
\end{align}
The Majorana $\gamma_D$ is situated on the ground island and stays decoupled from the rest of the system. The long-range Coulomb couplings $\Delta_{n,k}$ are
\begin{align}\nonumber
\Delta_{n,1}={}&U_{n,1}\frac{\cos\alpha_{n,25}}{\sqrt{\cos^2\alpha_{n,12}+\cos^2\alpha_{n,25}+\cos^2\alpha_{n,51}}}\\\nonumber
\Delta_{n,2}={}&U_{n,2}\frac{\cos\alpha_{n,34}}{\sqrt{\cos^2\alpha_{n,23}+\cos^2\alpha_{n,34}+\cos^2\alpha_{n,42}}}\\\nonumber
			&\times\frac{\cos\alpha_{n,51}}{\sqrt{\cos^2\alpha_{n,12}+\cos^2\alpha_{n,25}+\cos^2\alpha_{n,51}}}\\\nonumber
\Delta_{n,3}={}&U_{n,3}\frac{\cos\alpha_{n,42}}{\sqrt{\cos^2\alpha_{n,23}+\cos^2\alpha_{n,34}+\cos^2\alpha_{n,42}}}\\\nonumber
\Delta_{n,4}={}&U_{n,4}\frac{\cos\alpha_{n,23}}{\sqrt{\cos^2\alpha_{n,23}+\cos^2\alpha_{n,34}+\cos^2\alpha_{n,42}}}\\\nonumber
			&\times\frac{\cos\alpha_{n,g5}}{\sqrt{\cos^2\alpha_{n,4g}+\cos^2\alpha_{n,g5}+\cos^2\alpha_{n,54}}}\\\nonumber
\Delta_{n,5}={}&U_{n,5}\frac{\cos\alpha_{n,12}}{\sqrt{\cos^2\alpha_{n,12}+\cos^2\alpha_{n,25}+\cos^2\alpha_{n,51}}}\\\label{qubit_couplings}
			&\times\frac{\cos\alpha_{n,4g}}{\sqrt{\cos^2\alpha_{n,4g}+\cos^2\alpha_{n,g5}+\cos^2\alpha_{n,54}}}.
\end{align}
For computational purposes, one should be careful that the $\Delta_{n,k}$ do not change signs during the variation of the magnetic fluxes that takes place during a computational process. This may happen if some of the $\alpha_{n,kk'}$ in Eq. \eqref{phases} cross the value $\pi/2$. However, during any computation, maximally two of the fluxes are simultaneously turned on. Therefore, it is always possible to adapt the signs of the magnetic fluxes in such a way that the fluxes can be tuned in a range $|\Phi_{n,k}|=[0, \Phi_\textrm{max}]$, where $\Phi_\textrm{max} <h/4e$. We also notice that the signs of the couplings $\Delta_{n,k}$ in Eq.~(\ref{Hqubit}) depend on the signs of the microscopic tunnel couplings $E_M$. These signs will determine the chirality of the braiding of the Majorana fermions in each T-junction.

\subsection{Low-energy Hamiltonian in the readout configuration}\label{RAMMreadout}

During the readout, we set $\Phi_0=\Phi_\textrm{max}$ and all other fluxes $\Phi_{n,k}=0$. Following the same reasoning of Appendix~\ref{readout}, we set $\phi_{n,1}=\phi$ and $\phi_{n,k\neq 1}=0$ for each topological qubit. The Lagrangian for the {\sc ramm} becomes
\begin{align}\nonumber
\mathcal{L}=&\frac{\hbar}{8e^2}C\dot\phi^2+\frac{\hbar}{2e}\left(q_\textrm{tot}+\sum_{n=1}^N\, e\left(\tfrac{1}{2}-\tfrac{1}{2}\,i\gamma_{n,11}\gamma_{n,12}\right)\right)\dot\phi\\
					&-E_{J, 0}  (1-\cos\phi)-\sum_{n=1}^N\left.\Omega^\dagger_n V_M^{(n)}\Omega_n\right|_{\phi_{n,k}=0}
\end{align}
where $V_M^{(n)}$ describes the Majorana-Josephson potential for the three T-junctions in each topological qubit $n$,
\begin{equation}
\Omega_n=\prod_{k=1}^5\e^{i(1-i\gamma_{n,k1}\gamma_{n,k2})\phi_{k}/4}\label{Omega_n},
\end{equation}
\begin{align}
C&=C_0+\sum_{n=1}^N\sum_{k=2}^5 C_{B,k}+\sum_{n=1}^N C_{G,1}
\end{align}
and
\begin{equation}
q_\textrm{tot}=q_0+\sum_{n=1}^N\,q_{n,1}.
\end{equation}
The low-energy Hamiltonian of the system can now be derived analogously as in Appendix~\ref{readout}. By using the equality
\begin{align}
&\cos\left(\pi q_\textrm{tot}/e+\pi\sum_{n=1}^N\, \left(\tfrac{1}{2}-\tfrac{1}{2}\,i\gamma_{n,11}\gamma_{n,12}\right)\right)\nonumber \\ &=\prod_{n=1}^N\,i\gamma_{n,11}\gamma_{n,12}\,\cos\left(\pi q_\textrm{tot}/e\right)
\end{align}
we find\begin{align}\label{Hramm}
\tilde{H}_\textsc{ramm}&=\sigma_{z}\bigl[\tfrac{1}{2}\hbar\Omega_{0}+\mathcal{P}\,\Delta_{+}\cos(\pi q_\textrm{tot}/e)\bigr]\nonumber\\&\qquad+\mathcal{P}\,\Delta_{-}\cos(\pi q_\textrm{tot}/e)
\end{align}
where $\mathcal{P}$ is now the joint parity operator of the Majorana fermions at the measurement islands
\begin{equation}\label{Pmulti}
\mathcal{P}=\prod_{n=1}^N i\gamma_{n,F}\gamma_{n,E}.
\end{equation}

The couplings $\Delta_{\pm}$ decrease exponentially with the number of topological qubits involved in a single {\sc ramm} register
\begin{equation}
\Delta_\pm = \delta_{\pm} \prod_{n=1}^N \frac{\cos \alpha_{n,25}}{\sqrt{\cos^2\alpha_{n,12}+\cos^2 \alpha_{n,25}+\cos^2 \alpha_{n,51}}}. 
\end{equation}
In the design of a {\sc ramm} register, shown in Fig.~3b in the main text, the frequency shift $\omega_{\rm shift}$ is decreased by all topological qubits, including the ones which are not involved in a given multiqubit measurement. This limitation of {\sc ramm} can be relaxed in a more optimal design, where additional tunable Josephson junctions are introduced from the measurement island to the ground. In this case only the topological qubits involved in the given measurement contribute to the decrease of frequency shift. The expense one needs to pay for introducing new Josephson junctions is that the gauge invariant fluxes have more complicated magnetic flux dependence and several Josephson couplings need be simultaneously controlled when the Coulomb couplings are turned on. We point out that although we have explicitly considered the control of the Coulomb couplings with the help of magnetic fluxes, at least some of the macroscopic control parameters $E_{J,k}/E_{C,k}$ of the superconducting islands can alternatively be controlled with gates.

\section{Universal gates for quantum computation}\label{app_gates}

\begin{figure*}[tb]
\centerline{\includegraphics[width=0.7\textwidth]{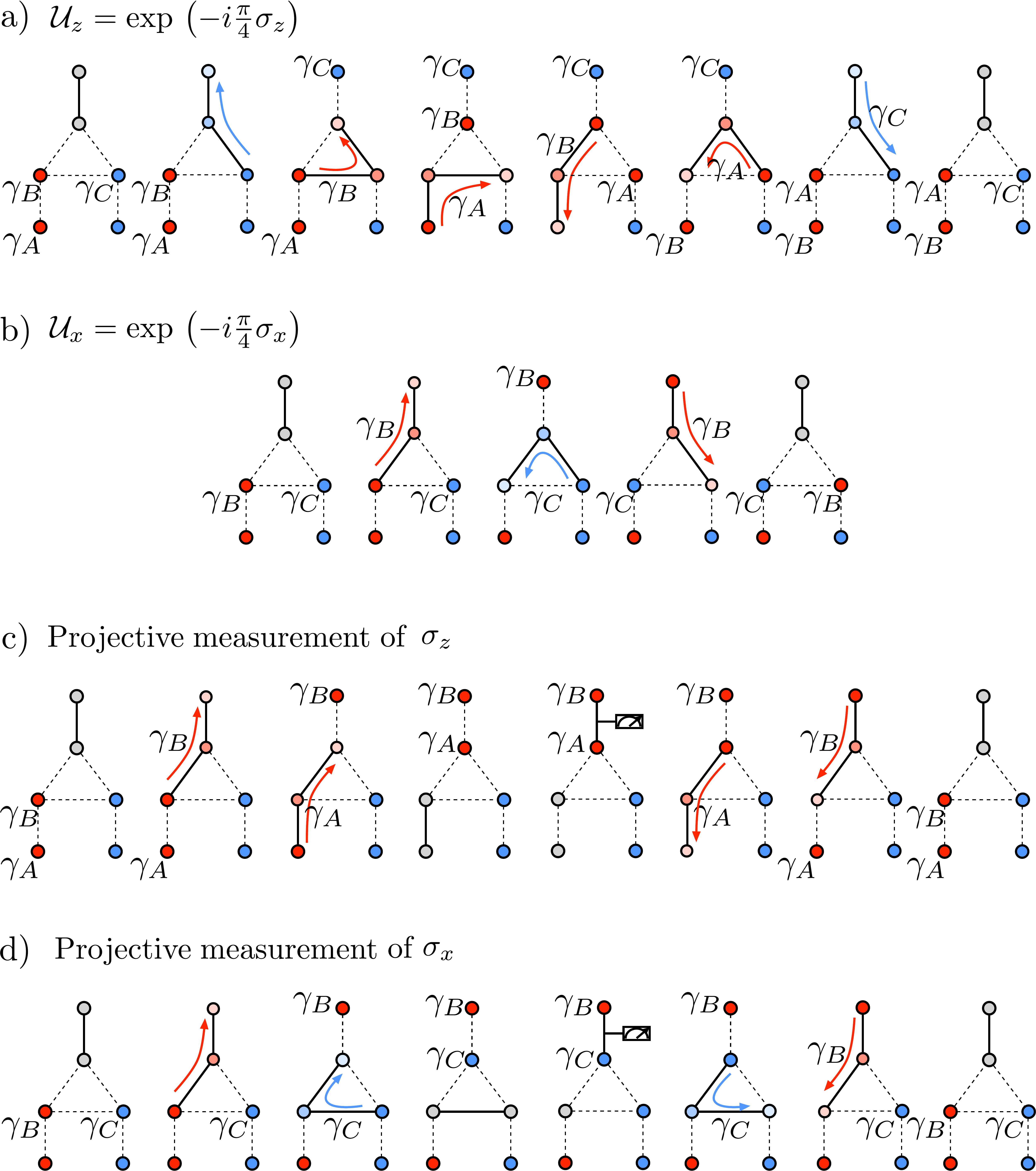}}
\caption{Flux-controlled sequences of operations that realize single-qubit Clifford gates and projective measurement on the Pauli basis.} \label{fig:1qubit_operations}
\end{figure*}

The {\sc ramm} setup allows us to perform universal quantum computation in a fault-tolerant way. To show this, it is necessary to implement a universal basis of quantum gates using only braiding operators and multi-qubit measurements as building blocks, thus ensuring the possibility of obtaining arbitrary multi-qubit gates with errors that are exponentially small in the macroscopically tunable parameters. One possible set of gates allowing for universal quantum computation are the single-qubit Clifford gates, the {\sc cnot} gate and the $\pi/8$  phase gate. In the following we explain how to realize these gates in a {\sc ramm} architecture.

\subsection{Notation}

Each topological qubit $n$ has four computational Majoranas $\gamma_{n,A},\gamma_{n,B},\gamma_{n,C},\gamma_{n,D}$ and two ancillary Majoranas $\gamma_{n,E},\gamma_{n,F}$, which are needed to move or braid the computational ones. The Pauli matrices for each qubit can be chosen as
\begin{subequations}
\begin{align}
\sigma_{n,z}&=i\gamma_{n,A}\gamma_{n,B}\\
\sigma_{n,x}&=i\gamma_{n,B}\gamma_{n,C}\\
\sigma_{n,y}&=i\gamma_{n,A}\gamma_{n,C}.
\end{align}
\end{subequations}

\subsection{Single-qubit operations}

Projective measurements on the Pauli basis and a set of Clifford gates can be obtained by manipulating the positions of the four computational Majorana fermions in the triangular loop geometry. The positions of the computational Majoranas $\gamma_{n,A}, \gamma_{n,B}, \gamma_{n,C}$ can be changed using the ancillary Majorana $\gamma_{n,E}, \gamma_{n,F}$, which remain strongly coupled throughout the process. The corresponding qubit transformation can be derived either by a direct computation of the non-Abelian Berry phase acquired by the ground state wave function of the Hamiltonian \eqref{Hqubit}, or by following the evolution of the Majorana operators in the Heisenberg picture, as explained in detail in Ref.~\cite{clarke2011,halperin2012}. 

Exchanging the positions of $\gamma_{n,A}, \gamma_{n,B}$ (as represented in Fig.\ \ref{fig:1qubit_operations}a) or $\gamma_{n,B}$ and $\gamma_{n,C}$ (Fig.\ \ref{fig:1qubit_operations}b) respectively yields the braiding gates
\begin{align}
\mathcal{U}_z&=\e^{-i\tfrac{\pi}{4}\sigma_z}\,,\\
\mathcal{U}_x&=\e^{-i\tfrac{\pi}{4}\sigma_x}\,.
\end{align}
The chirality of the braiding operations (i.e., the sign of the exponent in $\mathcal{U}_z, \mathcal{U}_x$) is determined by the signs of the couplings of the qubit Hamiltonian, Eq. \eqref{Hqubit}. Physically, the sign depends on the induced charges on the Majorana islands, the values of the fluxes and the signs of the microscopic tunnel couplings $\pm E_M$ at the T-junctions. Here, we have made a specific choice of chiralities. Another possibility of chiralities would not be harmful as long as they remain constant during the computation processes. 

A combination of these two operations yields the quantum gate corresponding to the braiding of $\gamma_A$ and $\gamma_C$,
\begin{equation}
\mathcal{U}_y=\mathcal{U}^\dagger_x\,\mathcal{U}_z\,\mathcal{U}_x=\e^{-i\tfrac{\pi}{4}\sigma_y}.
\end{equation}
When combined with the $\pi/8$ phase gate described in Appendix~\ref{phasegate}, these quantum gates are sufficient to realize any single-qubit rotation.

To realize projective measurements on $\sigma_{n,z}$ (or $\sigma_{n,x}$), we first need to bring the two Majorana fermions $\gamma_{n,A}, \gamma_{n,B}$ (or $\gamma_{n,B}, \gamma_{n,C}$) on the island connected to the bus, the one occupied by $\gamma_{n,E}, \gamma_{n,F}$ in Fig.\ \ref{fig:top_qubit}a. Then we measure the fermion parity operator \eqref{Pmulti}, where now the two Majoranas $\gamma_{n,E}, \gamma_{n,F}$ are replaced by the computational ones. For instance, in the case of a measurement of $\sigma_{n,z}$, we would measure the operator
\begin{equation}
\mathcal{P}=i\gamma_{n,A}\gamma_{n,B}\prod_{k\neq n} i\gamma_{k,E}\gamma_{k,F}\equiv\sigma_{n,z},
\end{equation}
since the parity of the ancillary Majorana of each topological qubit is preserved, $\mathcal{P}_{k,EF}=i\gamma_{k,E}\gamma_{k,F}=+1$. In the end, we bring the two computational Majoranas back to their original place. The whole operation, represented in Fig.\ \ref{fig:1qubit_operations}c and Fig.\ \ref{fig:1qubit_operations}d for $\sigma_{n,z}$ and $\sigma_{n,x}$ respectively, corresponds to the application of the projectors
\begin{subequations}
\begin{align}
\Pi_{z,n}(p)&=\tfrac{1}{2}\,\left(1+p \sigma_{n,z}\right),\\
\Pi_{x,n}(p)&=\tfrac{1}{2}\,\left(1+p \sigma_{n,x}\right)
\end{align}
\end{subequations}
to the wave function of the $N$ topological qubits. Here, $p=\pm 1$ is the outcome of the measurement. Finally, a projective measurement on $\sigma_{n,y}$ is obtained as
\begin{equation}
\Pi_{y,n}(p)=\tfrac{1}{2}\,\left(1+p \sigma_{n,y}\right)=\mathcal{U}^\dagger_x\,\Pi_{z,n}(p)\,\mathcal{U}_x\;.
\end{equation}
Multi-qubit measurements on the Pauli basis are a straightforward extension of these projective measurements where Majorana modes on different topological qubits are moved according to Fig.\ \ref{fig:1qubit_operations} to achieve the required basis.

\subsection{CNOT gate}

Bravyi and Kitaev have demonstrated how to realize the {\sc cnot} gate with an algorithm that is based on the following expansion \cite{bravyi2002,bravyi2006}:
\begin{widetext}
\begin{eqnarray}
\exp\left(i \frac{\pi}{4} \gamma_0 \gamma_1 \gamma_2 \gamma_3\right) |\psi \rangle&=& 2 e^{i \theta} \exp\left(\frac{\pi}{4} (1-p_1p_2) \gamma_0 \gamma_1 \right) \exp\left(\frac{\pi }{4}(1-p_1p_2)\gamma_2 \gamma_3 \right) \exp\left(- \frac{\pi}{4} p_2 \gamma_2 \gamma_5\right) \nonumber\\ && \times \frac{1}{2} (1+p_2 i \gamma_2 \gamma_4) \frac{1}{2}(1-p_1\gamma_0 \gamma_1 \gamma_3 \gamma_4) |\psi \rangle,  \label{BK}
\end{eqnarray}
\end{widetext}
where $\theta$ is an unimportant overall phase, $\gamma_i$ ($i=0,...,5$) are Majorana operators and $p_i= \pm 1$ are measurements outcomes. The Majoranas $\gamma_4$ and $\gamma_5$ are used as ancillas and the wave function is initialized in state $(\gamma_4+ i \gamma_5)|\psi \rangle=0$.
Importantly, the Bravyi-Kitaev {\sc cnot} algorithm is based only on measurements and braidings of Majorana fermions. However, as one can see from Eq.~(\ref{BK}), its implementation requires a pair of ancillary Majoranas that must be coupled to two computational Majoranas in the target qubit, but must initially be completely independent on them. Due to the parity constraint in each topological qubit, this is impossible in the {\sc ramm} setup unless we extend the qubit layout shown in Fig.~3a in the main text. Rather than modifying the {\sc ramm} setup to account for these new ancillas, we propose an alternative version of the {\sc cnot} gate, which involves three topological qubits. This alternative version of the {\sc cnot} gate can be implemented with the quantum circuit shown in Fig.~4a in the main text.

In this circuit the role of the first measurement, with result $p_1$, and of the gate $R_1$ is to initialize the third ancillary qubit in the state $\ket{0}_a$. After that, a {\sc cnot} gate with $q_1$ as a control and $q_2$ as a target gate is obtained as:
\begin{widetext}
\begin{equation}
 \frac{1}{2}\e^{i\frac{\pi}{4}p_2p_3\sigma_{1,z}}\e^{i\frac{\pi}{4}p_2p_3\sigma_{2,x}}\e^{-i\frac{\pi}{4}p_3\sigma_{a,x}}\left(1+p_3\sigma_{a,y} \right)\left(1+p_2\sigma_{1,z}\sigma_{2,x}\sigma_{a,x} \right)\ket{q_1,q_2,0}=\e^{i\theta} \ket{q_1,q_1\oplus q_2,0}.  
\end{equation}
In terms of Majorana operators, this way of representing the {\sc cnot} relies on the following equality
 \begin{multline}
   \exp\left( \frac{\pi}{4} \gamma_{1A} \gamma_{1B} \gamma_{2B} \gamma_{2C} \gamma_{3A} \gamma_{3B}\right) \ket{\psi}_{12}\ket{0}_a=
    2 e^{i \theta}
    \exp\left(-\frac{\pi}{4}(1+p_2 p_3)\gamma_{1A}\gamma_{1B} \right) 
      \exp\left(-\frac{\pi}{4} (1+p_2 p_3) \gamma_{2B}\gamma_{2C} \right) \\
    \times   
 \exp\left(\frac{\pi}{4}p_3\gamma_{3B}\gamma_{3C} \right) 
\frac{1}{2} (1+ip_3 \gamma_{3A} \gamma_{3C})  \frac{1}{2} (1- i p_2 \gamma_{1A} \gamma_{1B} \gamma_{2B} \gamma_{2C} \gamma_{3B} \gamma_{3C}) \ket{\psi}_{12}\ket{0}_a,
\end{multline}
\end{widetext}
which can be considered an extension of Kitaev and Bravyi result. In this case the applied projections are all on products of parity operators from different qubits, which can be reduced to the form \eqref{Pmulti} as explained above  (see Fig.\ \ref{fig:1qubit_operations}); all the other operators are braiding operators within single topological qubits. 

\subsection{$\bf \pi/8$ Phase Gate \label{phasegate}}

To complete the set of universal single-qubit gates we must implement the $\pi/8$ phase gate
\begin{equation}
 T=\begin{pmatrix} 1 & 0 \\ 0 & \e^{i\frac{\pi}{4}}\end{pmatrix},
\end{equation}
with an accuracy comparable to the other gates.

For this purpose the best techniques are based on distillation protocols \cite{bravyi2005}. The basic idea of the distillation procedure is the use of several noisy qubits to prepare one qubit in a particular state, $|A\rangle=\left( \ket{0} + \e^{i\pi/4} \ket{1}\right)/\sqrt{2}$. A single ancilla qubit  prepared in the state $\ket{A}$ is enough to implement  the $\pi/8$ gate using the circuit shown in Fig.~4b in the main text. 

The distillation protocol of Ref.~\cite{bravyi2005} for the state $\ket{A}$ assumes that it is possible to prepare several noisy copies of $\ket{A}$ with an average initial error $\epsilon_i < 0.14$. In the {\sc ramm} setup this can be achieved by coupling the Majorana fermions to break the ground state degeneracy \cite{hassler2011}. A single distillation step is performed starting from 15 noisy qubits. Neglecting the errors in all the Clifford gates and measurements of the distillation process, the error of the final state after one iteration is approximately
\begin{equation}
 \epsilon_{\sf dist} \approx 35 \epsilon_i^3 
\end{equation}
in the limit of small $\epsilon_i$. 

Since 14 stabilizer multi-qubit measurements and 15 {\sc cnot} gates are involved in the distillation-decoding procedure, the error in the $\pi/8$ gate is approximately an order of magnitude larger than the errors occurring in braiding or in a single multi-qubit measurement. Moreover, assuming an achievable initial error $\epsilon_i=0.01$ \cite{hassler2011} only a single distillation step involving 15 noisy ancillas is needed to achieve a final error of the same order of measurement and gate errors, estimated as $\Delta_{\rm min}/\Delta_{\rm max} \sim 10^{-5}$. If the initial errors are larger or the gate errors are smaller, more distillation steps and a larger number of ancillas are preferable. Given the amount of qubits required, it is realistic to imagine that the distillation procedure will take place in one (or several) dedicated {\sc ramm} registers, so that it can happen in parallel with all other computation processes. In this way, whenever a $\pi/8$ phase gate is needed in the computation, it will only be 
necessary to teleport the distilled state $\ket{A}$ from the distillation register to the computational one.

We also note that alternatively to the $\pi/8$ gate, the universality can also be obtained with the help of $\pi/12$ gate. This gate can be distilled with fewer noisy copies of the relevant state and a single distillation step also requires less multi-qubit measurements \cite{bravyi2005}.
Moreover, the distillation can be improved by exploiting more efficient error correction codes: for example in Ref.~\onlinecite{meier2012} a different procedure is proposed that enables to obtain two distilled states $\ket{A}$ out of 10 noisy ancillas, providing a better scaling and threshold for the initial errors. Finally we must mention that the distillation techniques in Ref.~\onlinecite{bravyi2005} require not only multi-qubit measurements and braiding gates, but also a non-unitary dephasing process. However, it was shown in Ref.~\onlinecite{jochym-oconnor2012} that the dephasing process is not necessarily needed for the convergence of the noisy states to a high-fidelity final state.  

\section{Computation of the error thresholds}\label{app_thresholds}

Multi-qubit measurements give significant advantages in quantum error correction, as compared to the usual schemes where only single- and two-qubit operations are available. The advantages obtained are twofold. Firstly, multi-qubit measurements allow to significantly increase error thresholds. Secondly, the overhead in computational resources can be substantially decreased. 

Quantum error correction schemes are generally based on measurements of multi-qubit operators, usually referred to as stabilizer generators $g_i$ \cite{nielsen2010}. Their outcomes give error syndromes, $\beta_i$, which uniquely characterize the errors and the qubits where they have occurred. Depending on the error correction scheme, a different number of errors can be corrected. 

For simplicity, we consider the Steane 7-qubit quantum code \cite{steane1996}, which encodes a logical qubit into seven physical qubits and can recover an arbitrary error occurring in any of the physical qubits. Its stabilizer generators are $g_1=X_1 X_5 X_6 X_7$, $g_2=X_2 X_4 X_6 X_7$, $g_3=X_3 X_4 X_5 X_6$, $g_4=Z_1 Z_3 Z_4 Z_7$, $g_5=Z_2 Z_3 Z_5 Z_7$, and $g_6=Z_1 Z_2 Z_3 Z_6$.  An error detected on the $i$-th qubit can be corrected by implementing a  $X_i$, $Z_i$ or $X_iZ_j$ gate, depending on the type of the error. 

\begin{figure}[t!]
\centerline{\includegraphics[width=0.9\linewidth]{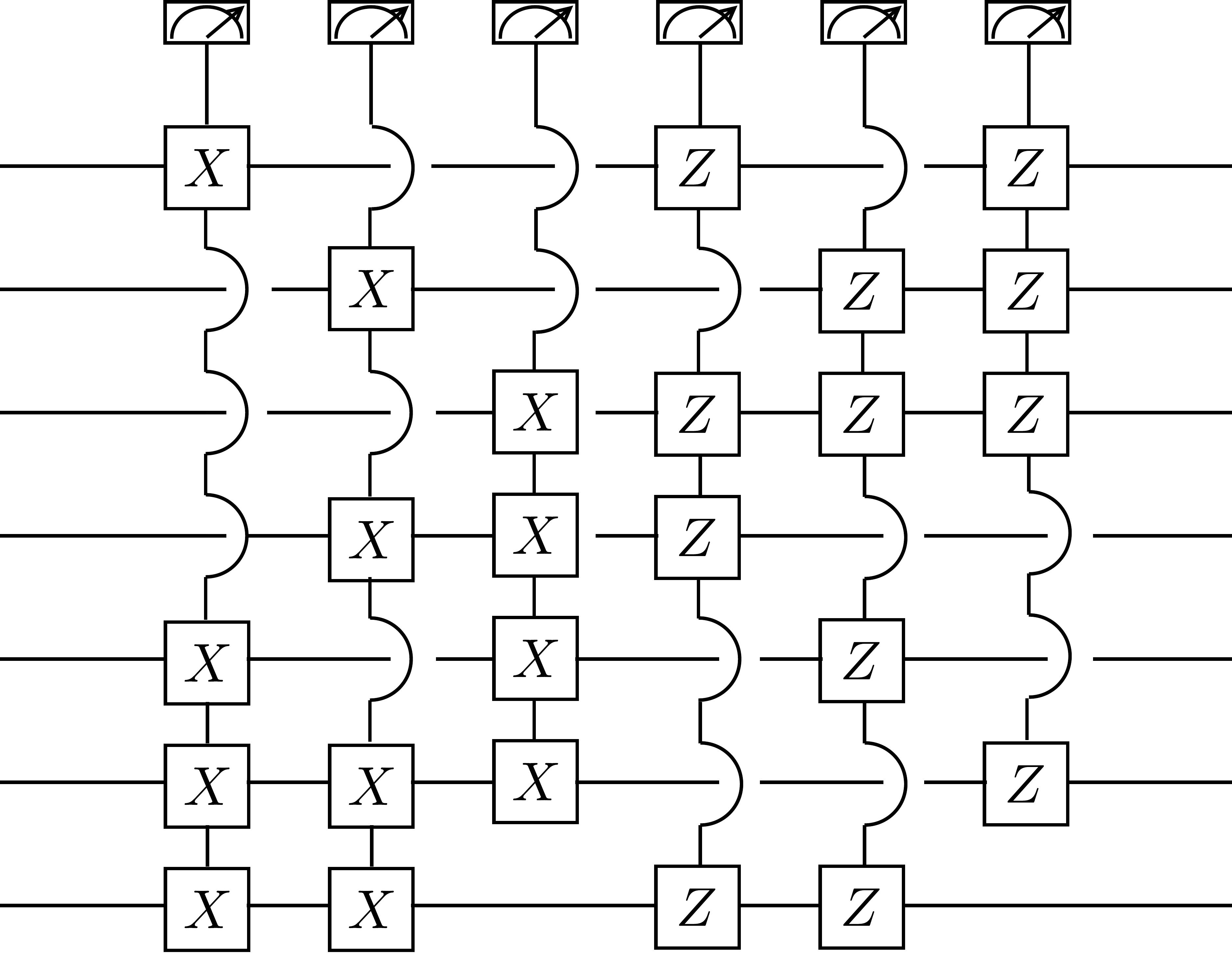}}
\caption{Measurement of the six generators of the Steane code. This circuit can be realized directly in a {\sc ramm} architecture.}
\label{fig:steane_ramm}
\end{figure}

In order to quantitatively compare the advantages obtained with the help of multi-qubit measurements to conventional schemes, we calculate the error threshold for a quantum memory. The error correction circuit consists of periodic syndrome measurements and recoveries, interrupted by a time-interval of $N$ time steps. Time steps are defined so that a single gate (or measurement) can be performed within one time step. Our error model consists of storage errors, gate errors, data errors during the measurement and errors in the measurement outcomes.  The corresponding error probabilities are $\epsilon_{\textrm{st}}$, $\epsilon_{\textrm{g}}$, $\epsilon_{\textrm{dm}}$, and $\epsilon_{\textrm{om}}$, respectively. All the errors are considered independent.  In order to obtain the error threshold, we need to calculate the probability of failure happening during a single period of the error correction circuit, assuming that no failure has happened before that point. To 
keep the calculation tractable, we assume that two errors in different qubits always result in failure (independently on the type of errors), and that this happens also when one of the errors occurs during the syndrome-recovery part of the circuit and the other error has happened earlier in the circuit.  Moreover, we assume that the errors occurring during the syndrome-recovery part of the circuit never get corrected by the same syndrome-recovery part of the circuit.  This way we obtain that the probability of failure during a single period of the circuit is:

\begin{widetext}
\begin{equation}\label{Pfailure}
\mathbb{P}(\textrm{failure}, N) \approx \mathbb{P}_{\textrm{om}}(2) + \mathbb{P}_{\textrm{om}}(1) \sum_i \big(2 \mathbb{P}_{i, \textrm{sr}}+\mathbb{P}_{i, \textrm{N}}\big) + \sum_{i<j} \bigg[ \big(2 \mathbb{P}_{i, \textrm{sr}}+\mathbb{P}_{i, \textrm{N}}\big) \big(2 \mathbb{P}_{j, \textrm{sr}}+\mathbb{P}_{j, \textrm{N}}\big) - \mathbb{P}_{i, \textrm{sr}} \mathbb{P}_{j, \textrm{sr}} \bigg].
\end{equation}
\end{widetext}

Here $\mathbb{P}_{\textrm{om}}(m)$ is the probability of having $m$ errors in the measurement outcomes, $\mathbb{P}_{i, \textrm{sr}}$ is the probability of obtaining single error in qubit $i$ during syndrome measurement and recovery, and $\mathbb{P}_{i, \textrm{N}}=N \epsilon_{\textrm{st}}$ is the probability of obtaining single error in qubit $i$ during the $N$ time steps between the successive error detections and recoveries.

\begin{figure*}[t!]
\centerline{\includegraphics[width=0.7\textwidth]{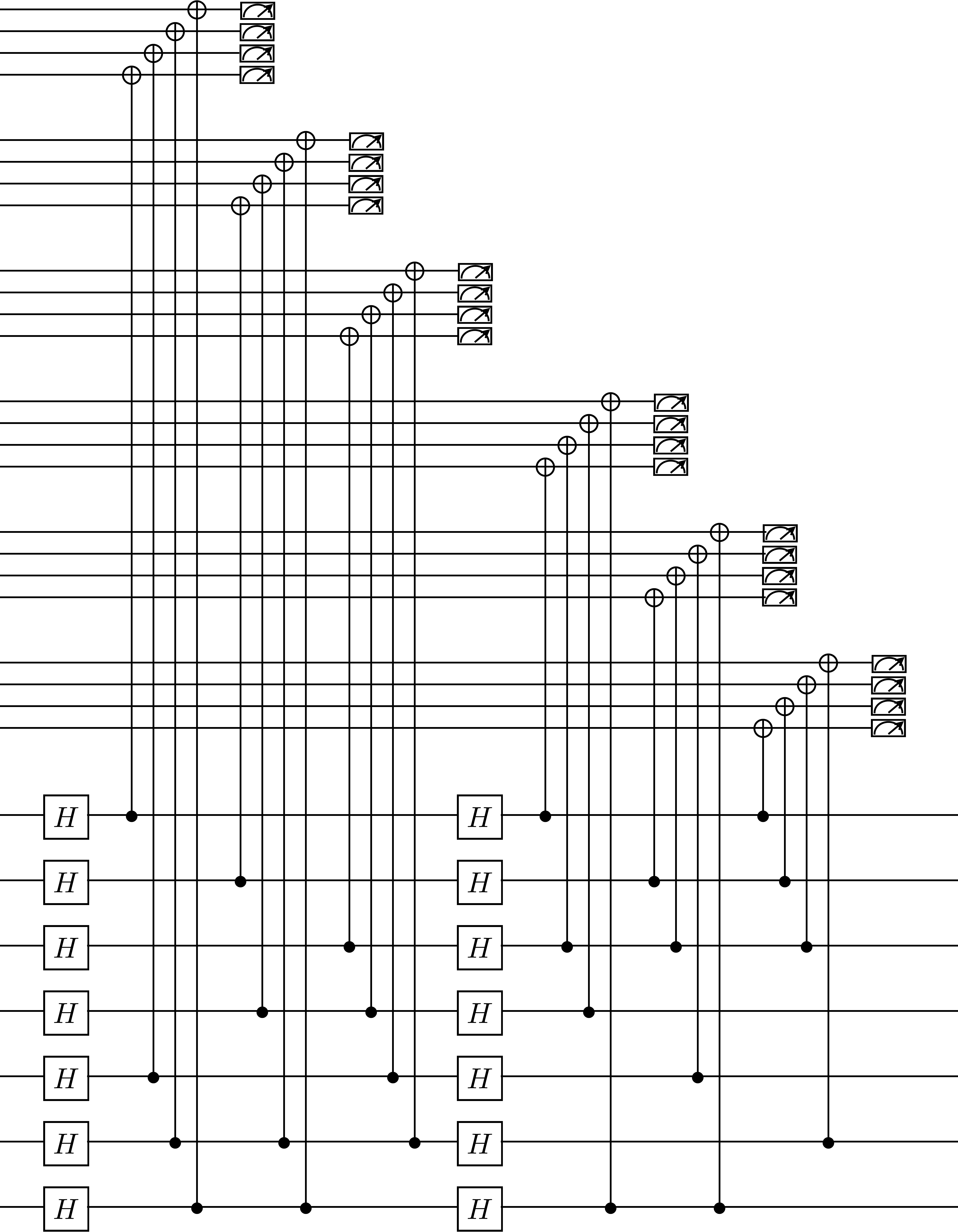}}
\caption{Quantum circuit to measure the generators of the Steane code in a traditional architecture that allows only for single- and two-qubit gates, and single-qubit measurements. Each of the six generator measurements is realized using four {\sc cnot} gates with an ancilla, which is in turn encoded using four physical qubits to avoid error propagation. This is the circuit we used to compare the error threshold with and without multi-qubit measurements.}
\label{fig:steane_circuit}
\end{figure*}

To estimate the error threshold we minimize the probability of failure per time step 
\begin{equation}
p_{\textrm{f}}=\min_{N>0}\{\mathbb{P}(\textrm{failure}, N)/(N+N_0) \},
\end{equation}
where $N_0$ is the number of time steps required to perform the syndrome measurements and the recovery.
The quantum error correction threshold is obtained by demanding that $p_{\textrm{f}}=\epsilon_{\textrm{st}}$. Because $p_{\textrm{f}} \propto \epsilon_{\textrm{st}}^2$, this equation determines a threshold value $\epsilon_{\textrm{st}}^{\textrm{th}}$. If $\epsilon_{\textrm{st}} < \epsilon_{\textrm{st}}^{\textrm{th}}$, the errors can be corrected by successively applying the scheme described above. For this kind of concatenated codes, the failure probability scales with the number of levels of encoding $k$ as 
\begin{equation}
p_{\textrm{f}, k}=\epsilon_{\textrm{st}}^{\textrm{th}} (\epsilon_{\textrm{st}}/\epsilon_{\textrm{st}}^{\textrm{th}})^{2^k},
\end{equation}
whereas the number of physical qubits needed to construct the logical qubits scales as $7^k$. In addition to the physical qubits needed for construction of the logical qubits, a large number of ancillas are typically needed to perform the syndrome measurements. These ancillas constitute the overhead in the required computational resources. 

\subsection{Realization of the Steane code with the {\sc ramm}}

In the case of the {\sc ramm}, the syndromes can be directly measured. For simplicity, we assume that one single-qubit gate is always performed during the recovery part of the circuit. Considering that each qubit is on average involved in $24/7$ measurements, the total number of time-steps required to perform the syndrome measurements is $6$, and the circuit contains $6$ measurements, we obtain 
\begin{subequations}\label{Pramm}
\begin{align}
\mathbb{P}_{\textrm{om}}(1)&=6\,\epsilon_{\textrm{om}}\,,\\
\mathbb{P}_{\textrm{om}}(2)&=\tfrac{1}{2}\cdot 6 \cdot 5 \, \epsilon_{\textrm{om}}^2=15\epsilon_{\textrm{om}}^2\,,\\
\mathbb{P}_{\textrm{sr}}&
=\tfrac{24}{7} \epsilon_{\textrm{dm}}+\tfrac{24}{7} \epsilon_{\textrm{st}}+\tfrac{1}{7}\epsilon_{\textrm{g}}\,.
\end{align}
\end{subequations}
These values allow to compute explicitly $\mathbb{P}(\textrm{failure}, N)$ for the {\sc ramm} via Eq. \eqref{Pfailure}.

\subsection{Steane's code without multi-qubit measurements}

We want to compare the error threshold in {\sc ramm} with a reference system, where multi-qubit measurements are not available. The syndrome measurements are then performed with the help of ancillas. In particular, the fault-tolerant realization of the six syndrome measurements requires a total of 24 ancillas, each quadruplet being used for measuring one of the syndromes \cite{preskill1998} (see Fig.~\ref{fig:steane_circuit}).

Each syndrome is measured by first initializing the ancilla quadruplet in a Shor state, which guarantees that measuring the four ancillas will not destroy the state encoded in the logical qubit. The second step consists of encoding the syndrome into the quadruplet, which requires performing a total of four {\sc cnot} gates between different ancillas and physical qubits. Since these involve independent qubit pairs, we assume that these four gates are performed simultaneously. Additionally, we assume that the syndrome is measured immediately after the {\sc cnot} gates and the initialization of the ancilla quadruplet takes place already before the syndrome measurements.
Because errors occurring in the ancillas essentially have the same effect as the errors in the measurement outcomes, we include all possible ancilla errors in $\mathbb{P}_\textrm{om}(m)$.

The initialization of the ancillas to a Shor state is explained in Ref.~\onlinecite{preskill1998}. It involves 7 time steps with 5 {\sc cnot} and 5 Hadamard gates. Moreover, a measurement is required to confirm that the Shor state was successfully encoded, otherwise the initialization process is repeated. We only consider gate and storage errors occurring in the initialization of the four ancillas. Each of the  ancillas is acted on with 13/4 gates on average.

The syndrome measurements involve 9 time steps and each of the 7 physical qubits is acted upon with $38/7$ gates on average, while recovery part only involves one single-qubit gate.  Finally, we need to take into account the errors occurring in any of the 24 ancillas during the syndrome block, which contribute to $\mathbb{P}_\textrm{om}$. This way we obtain 
\begin{subequations}\label{Pnormal_architecture}
\begin{align}
\mathbb{P}_\textrm{om}(1)&= 24\, (P_\textrm{init} +  P_\textrm{syndrome})\,, \\
\mathbb{P}_\textrm{om}(2)&= \tfrac{24\times 23}{2}\, (P_\textrm{init} + P_\textrm{syndrome})^2\,, \\
\mathbb{P}_\textrm{sr}&= \tfrac{38}{7}\epsilon_\textrm{g} + \tfrac{25}{7} \epsilon_\textrm{st} + \tfrac{1}{7}\epsilon_\textrm{g} + \tfrac{6}{7} \epsilon_\textrm{st}\,,
\end{align}
\end{subequations}
with
\begin{subequations}
\begin{align}
 P_\textrm{init} &= \epsilon_\textrm{om} +  \epsilon_\textrm{dm}\,, \\
 P_\textrm{syndrome} &= \tfrac{13}{4}\epsilon_\textrm{g} + \tfrac{15}{4} \epsilon_\textrm{st} + \epsilon_\textrm{g} + \epsilon_\textrm{om} + \tfrac{72}{24} \epsilon_\textrm{st}.
\end{align}
\end{subequations}
These values allow to compute $\mathbb{P}(\textrm{failure}, N)$ in the absence of multi-qubit measurements.

\begin{figure}[tb]
\centerline{\includegraphics[width=0.9\linewidth]{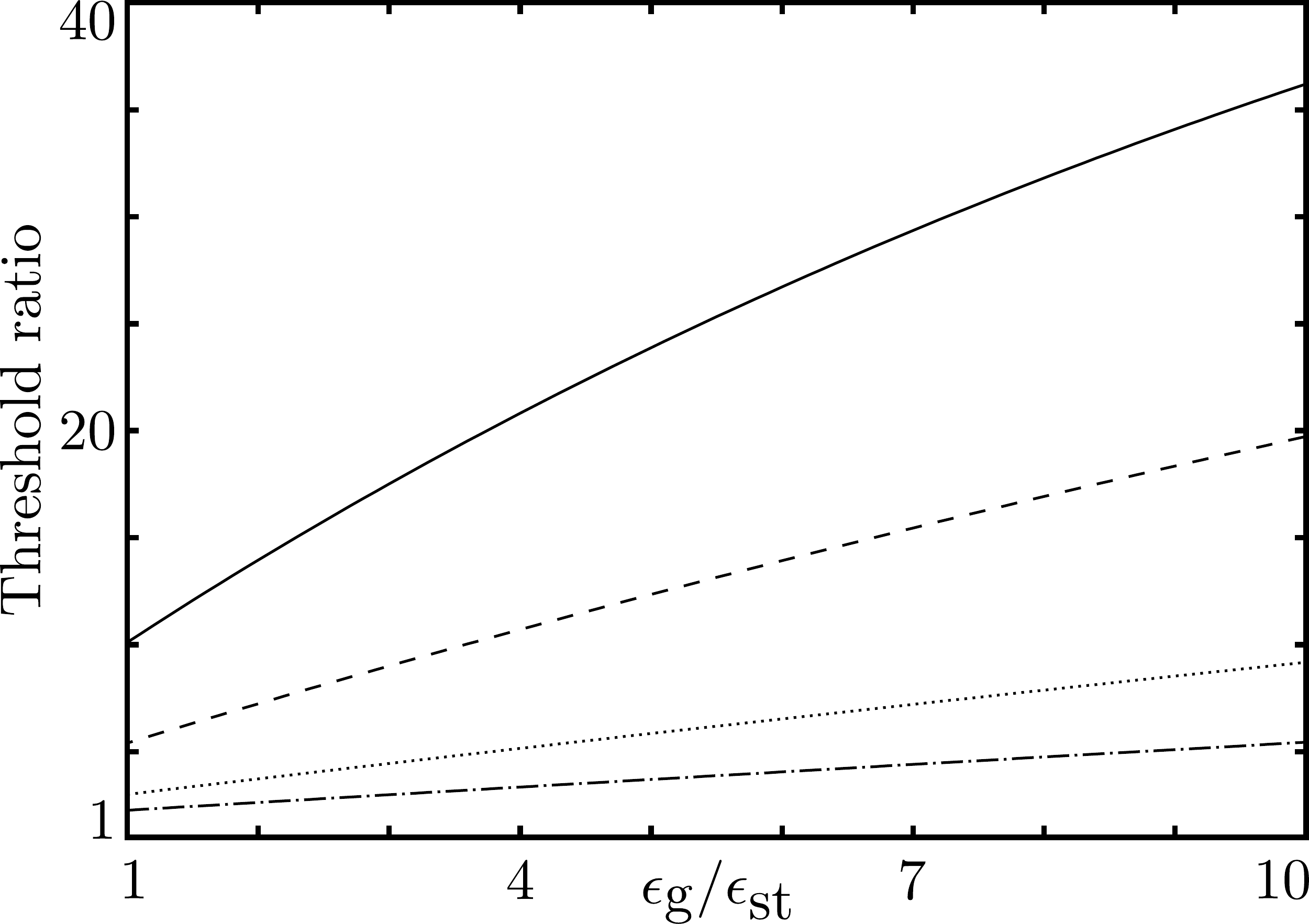}}
\caption{Ratio of the Steane code error thresholds with and without multi-qubit measurements as a function of the ratio between gate and storage errors, $\epsilon_\textrm{g}/\epsilon_\textrm{st}$. The solid, dashed, dotted, and dash-dotted curves correspond to ratios $\epsilon_\textrm{om} / \epsilon_\textrm{st} = \epsilon_\textrm{dm} / \epsilon_\textrm{st} = 1, 2, 5$, and $10$, respectively.}
\label{fig:threshold_ratio}
\end{figure}
 
\begin{figure}[tb]
\centerline{\includegraphics[width=0.9\linewidth]{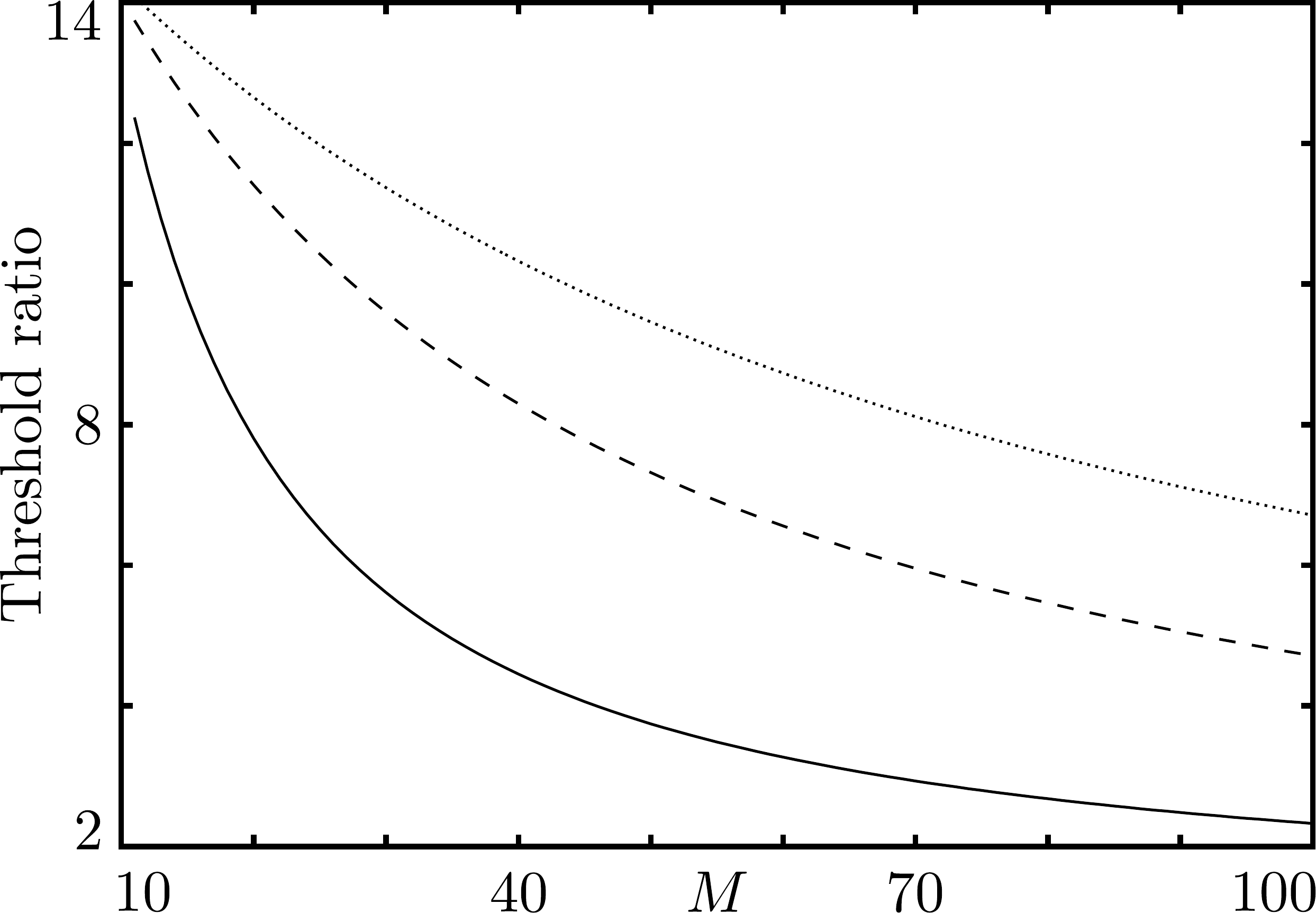}}
\caption{Ratio of the computational error thresholds with and without multi-qubit measurements as a function of $M$. Here $\epsilon = \epsilon_\textrm{om}=\epsilon_\textrm{dm}=\epsilon_\textrm{g}$, with $\epsilon / \epsilon_\textrm{st}=1$ (solid), 5 (dashed), and 10 (dotted). The range of $M$ starts from 11, because of the condition $M-N_0-1\geq0$.}
\label{fig:threshold_comp_ratio}
\end{figure}

\subsection{Comparison of the error thresholds for the quantum memory}

We minimize the probability of failure per time step with respect to $N$ for both implementations of the error correction scheme. We characterize the relative probabilities of errors by fixing the ratios $\epsilon_{\textrm{g}}/\epsilon_{\textrm{st}}$, $\epsilon_{\textrm{dm}}/\epsilon_{\textrm{st}}$ and $\epsilon_{\textrm{om}}/\epsilon_{\textrm{st}}$, and calculate the error threshold for $\epsilon_{\textrm{st}}$. 
 Results are shown in Fig.~\ref{fig:threshold_ratio}. 
 We find that for $\epsilon_{\textrm{g}}=\epsilon_{\textrm{dm}}=\epsilon_{\textrm{om}}=\epsilon_{\textrm{st}}$ the error threshold of the {\sc ramm} is approximately an order of magnitude larger than the error threshold of a reference architecture that can only perform single- and two-qubit operations. The ratio of the error thresholds for the different architectures becomes smaller with increasing measurement errors (larger ratios $\epsilon_{\textrm{dm}}/\epsilon_{\textrm{st}}$ and $\epsilon_{\textrm{om}}/\epsilon_{\textrm{st}}$), because it becomes favorable to increase the waiting time between the consequent error correction steps; but even for $\epsilon_{\textrm{g}}=\epsilon_{\textrm{dm}}=\epsilon_{\textrm{om}}=10 \epsilon_{\textrm{st}}$ we still find that the {\sc ramm} has an error threshold five times larger than the reference architecture. 
 
 \subsection{Comparison of the error threshold in quantum computation}
 
To estimate the error threshold in quantum computation, we consider an algorithm where each qubit participates in a two-qubit gate with a randomly chosen other qubit after every $M>N_0$ time steps. We assume that  the syndrome and recovery steps are performed once after each two-qubit gate. To estimate the error threshold we calculate the probability of failure in any one of the logical qubits during the $M$-step period. To keep the calculation tractable, we consider that all the errors appearing in a logical qubit during the syndrome and recovery steps just before the two-qubit gate propagate to the other qubit. Notice that due to the special construction of the Steane code, the  error occurring in $i$th physical qubit in one of the logical qubits will affect only the $i$th physical qubit in the other logical qubit. As before, we assume that two errors in a single logical qubit always result in failure. This way, we find

\begin{widetext}
\begin{eqnarray}
\mathbb{P}(\textrm{failure}, M) &\approx& \mathbb{P}_{\textrm{om}}(2) + \mathbb{P}_{\textrm{om}}(1) \sum_i \big(3 \mathbb{P}_{i, \textrm{sr}}+\epsilon_{\textrm{g}}+\mathbb{P}_{i, M-N_0-1}\big) \nonumber \\ &&+ \sum_{i<j} \bigg[ \big(3 \mathbb{P}_{i, \textrm{sr}}+\epsilon_{\textrm{g}}+\mathbb{P}_{i, M-N_0-1}\big) \big(3 \mathbb{P}_{j, \textrm{sr}}+\epsilon_{\textrm{g}}+\mathbb{P}_{j, M-N_0-1}\big)  - 2 \mathbb{P}_{i, \textrm{sr}} \mathbb{P}_{j, \textrm{sr}} \bigg]\,,
\end{eqnarray}
\end{widetext}
which we compute for both architectures using Eqs. \eqref{Pramm}, \eqref{Pnormal_architecture}. The probability of failure per time step is then
\begin{equation}
p_{\textrm{f}}=\mathbb{P}(\textrm{failure}, M)/M,
\end{equation}
and the threshold for quantum error correction can be determined by comparing this probability to the probability of failure  without error correction. Results are shown in Fig.~\ref{fig:threshold_comp_ratio}. Similarly as in the case of quantum memory, we find that the error threshold for performing the quantum computation can be an order of magnitude larger for the {\sc ramm}. 

\section{Characteristic energy scales of the problem}\label{app_energies}

We need to satisfy the following inequalities 
\begin{eqnarray}
E_{J,k}, \hbar \Omega_k, \Delta_{\rm g} & > & E_{J,0}, \hbar \Omega_0, \hbar \omega_0 \gg E_M, \Delta_{\rm max} \nonumber \\ &&  \gg k_B T, \Delta_{\rm min},
\end{eqnarray} 
where $\hbar \Omega_k \approx \sqrt{8 E_{J,k} E_{C,k}}$ is the plasma frequency of the small islands and $\Delta_{\rm g} \sim 100$ GHz is the induced gap in the nanowire. The condition $E_M, \Delta_{\rm max}   \gg k_B T$ is required to guarantee a relaxation to the ground state.
In the earlier sections we assumed that $E_M \gg U_k$ in order to turn our analytical calculations more transparent, but in view of the topological nature of the braiding our results  must remain valid also when $E_M$ and $\Delta_{\rm max}$ are comparable to each other. This is easy to understand, since independently on the ratio of $U_k$ and $E_M$ as long as the ground state manifold remains isolated from the excited states the adiabatic time-evolution operator for the braiding cycle takes the form of Eq. \eqref{Udef}, because of the topological nature of the operation.

Additionally, during the measurement  we need to satisfy the inequalities
\begin{equation}
E_M  \gg  \Delta_{+},
\end{equation}
and
\begin{equation}
\omega_{\rm shift} > \kappa, \label{resonatorineq}
\end{equation}
where $\kappa \sim 1-10$ MHz describes the characteristic cavity and qubit decay rates. The typical coupling between the microwaves and transmon is given by $g/2\pi \sim 100$ MHz.

\begin{figure}[tb]
\centerline{\includegraphics[width=0.98\linewidth]{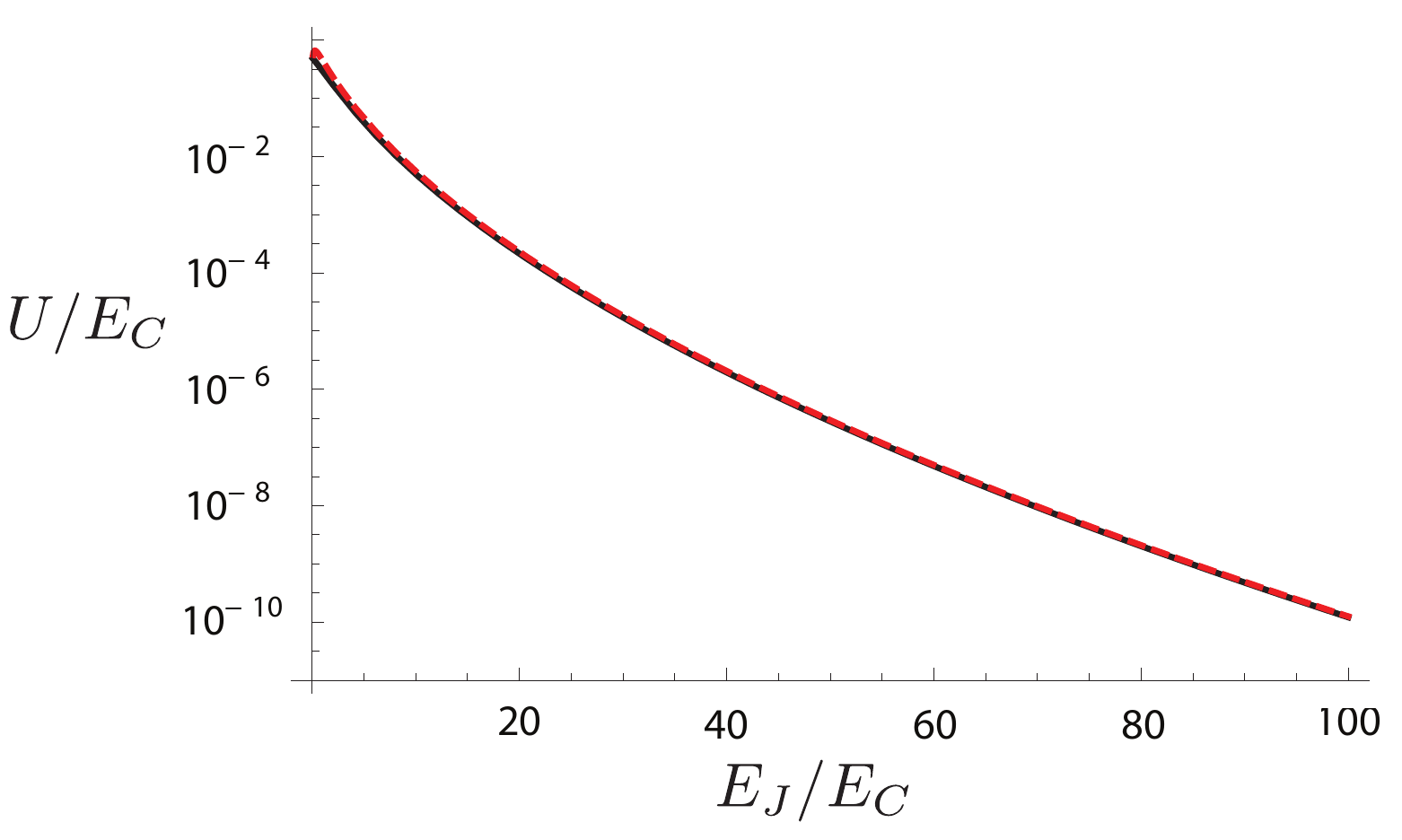}}
\caption{The dependence of the Coulomb coupling $U_k$ on the ratio $E_{J,k}/E_{C, k}$. The solid black line shows the exact solution obtained using Mathieu functions \cite{koch2007}, and the dashed red line shows the approximation [Eq.~(\ref{CoulombCoupApp})], which is valid in the asymptotic limit $E_{J,k}/E_{C, k} \gg 1$.}
\label{fig:numCoulombcoupling}
\end{figure}

The first set of inequalities can be satisfied with transmon parameters
$E_{J,0}, \hbar \Omega_0, \hbar \omega_0 \sim 100$ GHz, $E_M, \Delta_{\rm max} \sim 10$ GHz and $k_BT \sim 1$ GHz. The condition  $\Delta_{\rm max} \sim 10$ GHz  can be satisfied by having very large plasma frequency $\Omega_{k}$ or alternatively by tuning the $E_{J,k}(\Phi_{\rm max})/E_{C, k}$ ratio smaller than $10$, so that the superconducting islands do not stay in the transmon regime. As shown in Fig.~\ref{fig:numCoulombcoupling}, much larger Coulomb couplings can be achieved in this way, although  the asymptotic expression given by Eq.~(\ref{CoulombCoupApp}) is not valid anymore.

Importantly, the insensitivity of the couplings $\Delta_k$ to noise is needed only when the couplings are turned off. Since the topological protection of the braiding result only allows errors of order $\Delta_\textrm{min}/\Delta_\textrm{max}$, the exponential smallness of $\Delta_\textrm{min}$ guarantees that the result of the braiding cycle is not sensitive to low-frequency charge noise, which only affects the couplings which are turned on.

By assuming that $E_{J,0}/E_{C,0}=10$ during the measurement, we obtain, from Eq.~(\ref{transshift}), $\Delta_+ \sim 10^{-2} E_{J,0}$, which is consistent with the chain of inequalities. 
The  inequality  (\ref{resonatorineq})
  can be satisfied by tuning $\delta \omega$ and does not contradict with the requirement that we are working in the dispersive limit. 

As we have just remarked, the errors in the braiding are on the order $\Delta_{\rm min}/\Delta_{\rm max}$, which can be made exponentially small. The braiding and measurement should be performed fast in comparison to $\hbar/\Delta_{\rm{min}}$ and the characteristic quasiparticle tunneling time, which is on the order of milliseconds \cite{sun2012,riste2012}. In order that $\Delta_{\rm min}$ is limited by the charging energy, we need
$\Delta_{\rm g} \exp(-L/\xi)<\Delta_{\rm min}$, where $L$ is the length of the wire and $\xi$ is the Majorana decay length in the wire. Assuming that $\Delta_{\rm g} \sim E_{J,k}$, this means that $L \approx20 \xi$, so that $L$ should be at least several microns.

\bibliography{draft}

\end{document}